\documentclass[a4paper,fleqn]{cas-sc}

\usepackage[numbers,sort&compress]{natbib} 

\usepackage{revsymb}

\usepackage{algorithm}
\usepackage{algpseudocode}

\usepackage{graphicx}
\usepackage{subcaption}

\def\tsc#1{\csdef{#1}{\textsc{\lowercase{#1}}\xspace}}
\tsc{WGM}
\tsc{QE}
\tsc{EP}
\tsc{PMS}
\tsc{BEC}
\tsc{DE}

\begin{document}

\let\WriteBookmarks\relax
\def\floatpagepagefraction{1}
\def\textpagefraction{.001}
\shorttitle{Qprop-ND: A Nondipole Extension of the Qprop Package for Strong-Field Laser Physics}
\shortauthors{H. He \textit{et~al.}}

\title [mode = title]{Qprop-ND: A Nondipole Extension of the Qprop Package for Strong-Field  Laser  Physics}

\author[]{Haoyuan He}[orcid=0000-0000-0000-0000]
\ead{haoyuan@mpi-hd.mpg.de}
 \affiliation[]{organization={Max Planck Institute for Nuclear Physics},
            addressline={Saupfercheckweg 1}, 
            city={Heidelberg},
            postcode={69117}, 
                country={Germany}}

\author[]{Zhao-Han Zhang}[orcid=0000-0000-0000-0000]
                
\author[]{Karen Z. Hatsagortsyan}[orcid=0000-0002-1407-9122]

\author[]{Christoph H. Keitel}[orcid=0000-0002-1984-1470]

\begin{abstract}
The Qprop 3.0 package \cite{Bauer_2006,Mosert_2016,Tulsky_2020} for the solution of the time-dependent Schr\"odinger equation for an electron in a laser field and atomic potential is extended to incorporate nondipole effects related to the laser wave propagation. The developed code can be used for the simulation of strong field atomic phenomena in the nondipole regime, such as above-threshold ionization, ionization stabilization in high-frequency laser fields,  high-order harmonic generation, when the effects connected with the magnetic field are non-negligible or when the laser wavelength is comparable with the atomic size. Similar to the dipole Qprop  the wave function is expanded in spherical harmonics,  the time-propagation of the wave functions is performed using the Crank-Nicolson approach, and the photoelectron spectra are calculated employing  the t-SURFF or i-SURFV methods.  The validity and accuracy of the implementation are demonstrated through several illustrative examples.

\vspace{0.6em}

\noindent\textbf{Program summary}

\vspace{0.6em}

\noindent{\textit{Program title}: Qprop-ND}
\vspace{0.3em}

\noindent{\textit{Programming language}: C++}

\vspace{0.3em}

\noindent{\textit{Program Files doi}:} 

\vspace{0.3em}

\noindent{\textit{Operating system}: Linux(tested on Ubuntu 24.04.3 LTS)}

\vspace{0.3em}

\noindent{\textit{Licensing provisions}: GNU General Public License, version 3}

\vspace{0.3em}

\noindent{\textit{External routines/libraries}: GNU Scientific Library, Open MPI (optional)}

\vspace{0.3em}

\noindent{\textit{Nature of problem}: The nondipole effect becomes significant in the interaction of atoms with superintense laser fields or/and hard x-rays. An efficient computational framework is therefore essential to accurately predict the photoelectron momentum distributions in such regimes.}

\vspace{0.3em}

\noindent{\textit{Solution method}: The time-dependent Schrödinger equation, which incorporates a nondipole-corrected Hamiltonian, is solved by propagating the electronic wave function using the Crank-Nicolson method. The wave function is represented numerically by an expansion in spherical harmonics. }

\vspace{0.3em}

\noindent{\textit{Summary of revisions}: The nondipole laser field has been expanded with respect to the propagation term, yielding new terms in the Hamiltonian.}
\end{abstract}

\begin{keywords}
Qprop \sep i-SURFV \sep TDSE \sep Nondipole effect \sep Photoelectron spectra\sep Strong-field ionization \sep 
\end{keywords}

\maketitle

\section{Introduction}\label{sec:intro}

Since the invention of the chirped pulse amplification (CPA) technique  \cite{Strickland_1985,Strickland_2019,Mourou_2019}, laser fields  comparable to the inner-atomic Coulomb fields have been created, which led to the discovery of nonperturbative strong-field atomic phenomena such as above-threshold ionization (ATI), high-order harmonic generation (HHG), nonsequential double ionization,  stabilization, laser-induced electron diffraction, ultrafast holography \textit{etc.} \cite{Becker_2002,Agostini_2004,Becker_2012,Gavrila_2002,Protopapas_1997,Zuo_1996,Huismans_2011}, and  culminated by the emergence of attosecond science \cite{Corkum_2007,Krausz_2009,L'Huillier_2024,Agostini_2024,Krausz_2024}. The development of ultrastrong laser facilities, such as the Extreme Laser Infrastructure (ELI) \cite{ELI-ALPS,ELI-Beamlines,ELI-NP}, and others \cite{Vulcan10,PFS,AG,Apollon,OMEGA_EP,XCELS,SEL100PW} allowed to reach the relativistic regime of the laser-matter interaction with infrared (mid-infrared) lasers \cite{Maltsev_2003,Palaniyappan_2008,DiChiara_2008,Ekanayake_2013,Yandow_2024}.  Additionally, the development of the photoelectron measurement precision technique such as  the velocity map imaging spectrometer (VMIS) \cite{Eppink_1997,Weger_2013}, and the Cold Target Recoil Ion Momentum Spectroscopy
(COLTRIMS) or the so-called reaction microscope  \cite{Doerner_2000,COLTRIM}, allowed recently for observation of subtle  nondipole effects at a rather weak nonrelativistic laser intensities \cite{Ludwig_2014,Maurer_2018,Willenberg_2019,Hartung_2019,Haram_2019,Grundmann_2020,Lin_2021,
Hartung_2021,Lin_2022a,Mao_2025}.

In another front, with the advent of powerful free-electron laser facilities, such as the European x-ray free-electron lasers (XFEL) in Hamburg, the Linac Coherent Light Source (LCLS) at Stanford,  China’s upcoming soft x-ray free-electron lasers (SXFEL) in Shanghai and others, strong laser fields have been reachable in the high-frequency domain, which allowed for the observation of nonlinear ionization processes with XUV and x-ray light \cite{Rohringer_2012,Young_2018,Linker_2025}. The upgrades of these facilities promise more intense x-rays with shorter wavelengths. This may allow to enter nonperturbative and nondipole regimes of x-ray atom interaction, which have not yet been systematically explored and are expected to give rise to a range of intriguing phenomena. The nondipoleness here arises for two reasons, when the electron velocity in the field approaches the speed of light, or when the laser wavelength  becomes comparable with the atomic size, and in the strong hard x-ray regime these nondipole effects can be intertwined.

In the nonrelativistic and nonperturbative regime, the time-dependent Schr\"odinger equation (TDSE) provides a rigorous framework for accurately predicting and explaining the underlying dynamics of these phenomena. In the fully relativistic regime, the Dirac equation should be solved with corresponding numerical challenges, see e.g. \cite{Braun_1999,Telnov_2020,Mocken_2008,Selsto_2009,Vanne_2012,Ivanov_2015,Bauke_2017,Vembe_2024}. However, in the nondipole regime, the extension of TDSE including the laser wave propagation effects, is sufficient for the description of strong-field phenomena and can provide a detailed and quantitative understanding of these processes, see e.g. \cite{Forre_2005,Meharg_2005,Forre_2006,Dimitrovski_2009,Dondera_2012,Forre_2014,Ivanov_2016,Brennecke_2018,Ni_2020}.
 
Since its initial release  \cite{Bauer_2006}, the \textsc{Qprop} package has evolved into one of the powerful and widely used tools for solving the three-dimensional TDSE for a single active electron initially bound in a spherically symmetric potential and interacting with a laser field. In \textsc{Qprop}, the total wave function is expanded in spherical harmonics $Y_{\ell}^m(\Omega)$, which reduces the three-dimensional TDSE to a set of coupled one-dimensional radial equations for the partial-wave components $\phi_{\ell m}(r,t)$. These components are propagated in time using the Crank--Nicolson scheme, ensuring numerical stability and norm conservation even for long propagation times and intense laser fields.

For accurate and efficient simulations, a well-known difficulty arises from the fact that electrons easily acquire high momenta in strong laser fields, which necessitates large spatial grids and makes the calculation of the photoelectron momentum distribution (PMD) computationally demanding. This challenge is addressed in \textsc{Qprop} through the incorporation of the time-dependent surface flux method (t-SURFF) \cite{Mosert_2016} and its improved version, i-SURFV \cite{Tulsky_2020}. 
In the t-SURFF  the momentum information of ionized electrons is extracted from the time-dependent probability flux through a selected surface, thereby avoiding the need to propagate the spatially extended wave function over large distances. 
However, t-SURFF requires that the slowest electrons of interest reach the detection surface within the total simulation time, necessitating extended propagation - particularly for ultrashort laser pulses. The i-SURFV method overcomes this limitation by recognizing that, once the laser field is switched off, the system evolves under a time-independent Hamiltonian. This allows the post-pulse flux contribution, extending to infinity, to be evaluated in a single, efficient computation. With these advancements, \textsc{Qprop} provides a robust framework for investigating strong-field and nonperturbative laser matter interactions. 

In \textsc{Qprop}, the dipole approximation is employed, which assumes that the spatial dependence of the laser field can be neglected over the extent of the electronic wave function. While this approximation has been remarkably successful in describing most strong-field phenomena, its validity breaks down in the regime of superintense and ultrashort-wavelength laser fields, where both the spatial dependence of the electromagnetic field and the associated magnetic component can no longer be ignored.
The nondipole effects arising  from the magnetic-field-induced Lorentz force becomes important when the laser  strong field parameter is large $a_0=E_0/c\omega\gtrsim 0.1$, with the laser field amplitude $E_0$, and the frequency $\omega$, and the speed of light $c$  (atomic units are applied), or when the retardation effect of the electromagnetic wave is not negligible while the laser wavelength $\lambda$ becomes comparable with the atomic size $a_s$.
The nondipole effects introduce significant corrections to electron dynamics. It is known that in infrared laser fields the laser magnetic field will induce a momentum shift along the laser propagation direction and corresponding characteristic asymmetries in the angular distribution   \cite{Klaiber_2005}, while the peak of the PMD is shifted in the laser counterpropagating direction due to the interplay between the magnetic field and Coulomb attraction effects \cite{Ludwig_2014,Maurer_2018,Willenberg_2019}. Due to nondipole effects, the ponderomotive potential in the laser field becomes $p_z$-momentum dependent (along the laser propagation direction), which leads to a shift of ATI peaks in the opposite direction \cite{Lin_2022a,Mao_2025}. The under-the-barrier effect of the laser magnetic field induces a momentum shift of the photoelectron at the tunnel exit by an amount $I_p/3c$, where $I_p$ is the ionization potential \cite{Klaiber_2013,Klaiber_2022}. This shift leaves a clear signature in the PMD \cite{Hartung_2019} and modifies the simple-man picture of photon momentum sharing between the photoelectron and the ion \cite{Cricchio_2015}.
The magnetic field induced drift can suppress recollisions \cite{Dammasch_2001}, and consequently, HHG \cite{Keitel_1995,Walser_2000a,Milosevic_2000,Kylstra_2001,Chirila_2004}, and hinders the HHG extension above  10 keV photon energies \cite{Palaniyappan_2006,Hatsagortsyan_2008,Kohler_2011a,Kohler_adv,Klaiber_2017}.

In high-frequency laser fields $\omega\gtrsim I_p$, the magnetically induced drift can suppress the ionization stabilization effect \cite{Protopapas_1997,Telnov_2020,Latinne_1994,VazquezdeAldana_1999,Kylstra_2000,Telnov_Chu_2021}. A remarkable signature of nondipole effects at high-frequencies is the emergence of a distinct  three-lobe low-energy structure  in the angularly resolved PMD observed under high-intensity laser fields \cite{Forre_2006,Zhou_2013,Telnov_2020,Telnov_Chu_2021,Geng_2021}. This structure originates from the interplay between the Lorentz-force-induced drift of the electron in the laser propagation direction and the Coulomb interaction with the parent ion. The magnetic drift shifts the electron wave packet during the interaction in the laser propagation direction, which creates a net Coulomb momentum transfer in the opposite direction. Together, they break the forward--backward symmetry of PMD predicted by the dipole approximation. The two of the observed three lobes stem from the dipole regime angular distribution shifted backwards due to the mentioned asymmetry, while the third lobe -- exactly in the counterpropagation direction -- is a fully nondipole effect, created by the electron trajectories which are captured in the bound state in the dipole regime. This low-energy interference structure in PMD \cite{Geng_2021}, typically oriented opposite to the laser propagation direction, serves as a signature for the deviation of nondipole dynamics from the predictions of the dipole approximation in strong-field ionization in high-frequency fields. The accurate reproduction of this interference structure thus provides a stringent benchmark for theoretical models that incorporate nondipole effects.

In this work the \textsc{Qprop} 3.0 package is extended beyond the dipole approximation by including the corresponding nondipole terms in the Hamiltonian, thereby enabling detailed investigations of strong-field ionization dynamics in strong electromagnetic waves.

This paper is structured as follows. Sec.~\ref{sec:theory} presents the mathematical formulation of the nondipole effects. Sec.~\ref{sec:rel} describes the corresponding modifications to the \textsc{Qprop} algorithm. Sec.~ \ref{sec:examples} provides examples for strong-field ionization of hydrogen and helium in the nondipole regime, demonstrating the performance of the extended framework of \textsc{Qprop-ND}. Our conclusions are given in Sec.~\ref{concl}.

Unless otherwise stated, atomic units ($\hbar = |e| = m_e = 4\pi\epsilon_0 = 1$) are used throughout this work.


\section{Theoretical description of nondipole effects}\label{sec:theory}
\subsection{Nondipole correction of the Hamiltonian}
\noindent
We consider a nonrelativistic system with a single active electron, initially bound by a spherically symmetric potential of the atomic core, which evolves under the influence of an intense and  short wavelength laser field. The laser field is elliptically polarized in the $(x,y)$ plane and propagates along the $z$ axis, with the vector potential given by
\begin{equation}
    \mathbf{A}(z, t) = A_x(t-z/c)\,\hat{\mathbf{e}}_x + A_y(t-z/c)\,\hat{\mathbf{e}}_y.
\end{equation}
In the nondipole weakly relativistic regime of what in this paper, as described in the Introduction, the vector potential $\mathbf{A}(z,t)$ can be expanded to first order in $z/c$ as
\begin{equation}
\begin{aligned}
    \mathbf{A}(z,t)= \mathbf{A}(t) + \frac{z}{c}\,\mathbf{E}(t),
\end{aligned}
\end{equation}
where we have used the relation $\mathbf{E}(t) = -\partial_t \mathbf{A}(t)$.
The dynamics of the electron is described by the following TDSE:
\begin{equation}
    i\,\frac{\partial}{\partial t}\,|\Psi(t)\rangle = \hat{H}(t)\,|\Psi(t)\rangle,
\end{equation}
with the minimal-coupling Hamiltonian, including first-order nondipole corrections, given by
\begin{align}
    \hat{H}(t)
    &= -\frac{1}{2}\nabla^{2} 
    + U(\mathbf{r}) 
    - i\,V_{\text{im}}(\mathbf{r})
    \underbrace{- i\,A_{x}(t)\frac{\partial}{\partial x}
    - i\,A_{y}(t)\frac{\partial}{\partial y}}_{\hat{H}^{(int)}_{dipole} }\nonumber \\
    &\quad
    \underbrace{- i\,\frac{z}{c}\,\left[
        E_{x}(t)\frac{\partial}{\partial x}
        + E_{y}(t)\frac{\partial}{\partial y}
    \right]
    + \frac{z}{c}\,\left[
        A_{x}(t)E_{x}(t)
        + A_{y}(t)E_{y}(t)
    \right]}_{\hat{H}^{(int)}_{nondipole}},
\end{align}
where $U(\mathbf{r})$ is the atomic binding potential. 
The purely time-dependent term proportional to $A^{2}(t)$, 
arising from the minimal-coupling form 
$(\hat{\mathbf{p}} + \mathbf{A}(t))^{2}/2$, 
has been removed via the gauge transformation, 
while the nondipole term proportional to $E^{2}(t)$ of the order of $1/c^{2}$ is neglected in the considered weakly relativistic parameter regime. The  imaginary absorbing potential $V_{\mathrm{im}}(\mathbf{r})$  is introduced to prevent unphysical reflections of the wave function at the numerical boundaries. 

\subsection{Separation of the radial and angular variables in the TDSE}

The wave function $\Psi(\mathbf{r},t)$ is expanded in spherical harmonics 
$Y_{\ell}^{m}(\Omega)$ as
\begin{equation}
\Psi(\mathbf{r},t) = \frac{1}{r} 
\sum_{\ell=0}^{\infty} 
\sum_{m=-\ell}^{\ell} 
\Phi_{\ell m}(r,t) 
Y_{\ell}^{m}(\Omega),
\end{equation}
where $\Phi_{\ell m}(r,t)$ are the radial wave function components associated 
with each angular momentum channel $(\ell, m)$, and 
$\Omega$ is the solid angle defined by 
$\mathrm{d}\Omega_r = \sin\theta_r\, \mathrm{d}\theta_r\, \mathrm{d}\varphi_r$.
The TDSE for the radial part of the wave function with first-order nondipole corrections reads
\begin{equation}
\begin{aligned}
i \partial_t \Phi_{\ell m} 
&= \left( -\tfrac{1}{2} \frac{\partial^2}{\partial r^2} + V_\ell^{\text{eff}}(r) \right) \Phi_{\ell m} \\[4pt]
&\quad - \frac{ir}{2} \sum_{\ell' m'} \langle \ell m | ( e^{i\varphi} \tilde{A}^* + e^{-i\varphi} \tilde{A} ) \sin\theta | \ell' m' \rangle \, \partial_r\frac{1}{r} \Phi_{\ell' m'} \\[4pt]
&\quad - \frac{i}{2r} \sum_{\ell' m'} \langle \ell m | \tilde{A}^* e^{i\varphi} ( \cos\theta \, \partial_\theta + \tfrac{i}{\sin\theta} \partial_\varphi ) | \ell' m' \rangle \Phi_{\ell' m'} \\[4pt]
&\quad - \frac{i}{2r} \sum_{\ell' m'} \langle \ell m | \tilde{A} e^{-i\varphi} ( \cos\theta \, \partial_\theta - \tfrac{i}{\sin\theta} \partial_\varphi ) | \ell' m' \rangle \Phi_{\ell' m'} \\[4pt]
&\quad - \frac{i r^2}{2c} \sum_{\ell' m'} \langle \ell m | ( e^{i\varphi} \tilde{E}^* + e^{-i\varphi} \tilde{E} ) \cos\theta \, \sin\theta | \ell' m' \rangle \, \partial_r \tfrac{1}{r} \Phi_{\ell' m'} \\[4pt]
&\quad - \frac{i}{2c} \sum_{\ell' m'} \langle \ell m | \tilde{E}^* e^{i\varphi} \cos\theta \, ( \cos\theta \, \partial_\theta + \tfrac{i}{\sin\theta} \partial_\varphi ) | \ell' m' \rangle \Phi_{\ell' m'} \\[4pt]
&\quad - \frac{i}{2c} \sum_{\ell' m'} \langle \ell m | \tilde{E} e^{-i\varphi} \cos\theta \, ( \cos\theta \, \partial_\theta - \tfrac{i}{\sin\theta} \partial_\varphi ) | \ell' m' \rangle \Phi_{\ell' m'} \\[4pt]
&\quad + \frac{r}{c} \, \tilde{A} \tilde{E}^* \sum_{\ell' m'} \langle \ell m | \cos\theta | \ell' m' \rangle \Phi_{\ell' m'}
\end{aligned}
\end{equation}
where $V_\ell^{\text{eff}}(r)=U(r)-iV_{im}(r)+\frac{\ell (\ell +1)}{r^2}$ is the effective potential, $|{\ell m}\rangle  = |{Y_{\ell}^{m}}\rangle$, $\tilde{E} = E_x + i E_y$ and $\tilde{A} = A_x + i A_y$.
The ladder operators $\hat{L}_{\pm}$ are defined as
\begin{equation}
\hat{L}_{\pm} = -\frac{1}{\sqrt{2}}\, e^{\pm i\varphi} \left( \partial_\theta \pm i \cot\theta\, \partial_\varphi \right),
\end{equation}
which act on a spherical harmonic according to $\hat{L}_{\pm}|{\ell m}\rangle = \mp N_{\ell m}^{\pm} |{\ell m \pm 1}\rangle$,
with
\begin{equation}
N_{\ell m}^{\pm} 
= \sqrt{ \frac{ \ell(\ell+1) - m(m \pm 1) }{2} }
= \sqrt{ \frac{ (\ell \mp m)(\ell \pm m + 1) }{2} }.
\end{equation}
The radial operator $r\partial/\partial r$ in the nondipole terms of the Hamiltonian is represented as
\begin{equation}
r\frac{\partial}{\partial r} = \sqrt{r} \frac{\partial}{\partial r} \sqrt{r} - \frac{1}{2}
\end{equation}
where the term $\sqrt{r} (\partial/\partial r) \sqrt{r}$ is antisymmetric in radial space, thereby ensuring the Hermiticity of the resulting Hamiltonian. 
With the ladder operators and this decomposition, the TDSE can thus be written as
\begin{equation}
\begin{aligned}
i \partial_t \Phi_{\ell m} 
&= \left[ -\tfrac{1}{2} \frac{\partial^2}{\partial r^2} + V_\ell^{\text{eff}}(r) \right] \Phi_{\ell m} \\[6pt]
&\quad + i \sqrt{\tfrac{2\pi}{3}} \sum_{\ell' m'}
\Biggl\{
 \tilde{A}^* \langle \ell m | Y_1^{1} | \ell' m' \rangle \, \partial_r
 - \tilde{A} \langle \ell m | Y_1^{-1} | \ell' m' \rangle \, \partial_r \\[4pt]
&\qquad
 - \frac{\tilde{A}^*}{r} \langle \ell m | Y_1^{1} | \ell' m' \rangle (1 + m')
 + \frac{\tilde{A}}{r} \langle \ell m | Y_1^{-1} | \ell' m' \rangle (1 - m') \\[4pt]
&\qquad
 - \frac{\tilde{A}^*}{r} \langle \ell m | Y_1^{0} | \ell' m' + 1 \rangle N_{\ell' m'}^{+}
 + \frac{\tilde{A}}{r} \langle \ell m | Y_1^{0} | \ell' m' - 1 \rangle N_{\ell' m'}^{-}
\Biggr\} \Phi_{\ell' m'} \\[6pt]
&\quad + i \frac{2 \sqrt{2}\pi}{3c} \sum_{\ell' m'}
\Biggl\{
 \tilde{E}^* \langle \ell m | Y_1^{0} Y_1^{1} | \ell' m' \rangle \sqrt{r} \partial_r \sqrt{r}
 - \tilde{E} \langle \ell m | Y_1^{0} Y_1^{-1} | \ell' m' \rangle \sqrt{r} \partial_r  \sqrt{r}\\[4pt]
&\qquad
 - \tilde{E}^{*} \langle \ell m | Y_1^{0} Y_1^{1} | \ell' m' \rangle (1 + m')
 + \tilde{E} \langle \ell m | Y_1^{0} Y_1^{-1} | \ell' m' \rangle (1 - m') \\[4pt]
&\qquad
 - \tilde{E}^{*} \langle \ell m | Y_1^{0} Y_1^{0} | \ell' m' + 1 \rangle N_{\ell' m'}^{+}
 + \tilde{E} \langle \ell m | Y_1^{0} Y_1^{0} | \ell' m' - 1 \rangle N_{\ell' m'}^{-}
\Biggr\} \Phi_{\ell' m'} \\[6pt]
&\quad + i \frac{\sqrt{2}\pi}{c} \sum_{\ell' m'}
\Bigl(
 \tilde{E} \langle \ell m | Y_1^{0} Y_1^{-1} | \ell' m' \rangle
 - \tilde{E}^{*} \langle \ell m | Y_1^{0} Y_1^{1} | \ell' m' \rangle 
\Bigr) \Phi_{\ell' m'}  \\[4pt]
&\quad + \sqrt{\tfrac{4\pi}{3}} \frac{r}{c} \tilde{A} \tilde{E}^{*}
 \sum_{\ell' m'} \langle \ell m | Y_1^{0} | \ell' m' \rangle \Phi_{\ell' m'} .
\end{aligned}
\label{lm}
\end{equation}
The calculation of dipole terms is already given in Ref.~\cite{Bauer_2006}, therefore, we only present the results of nondipole coefficients in this paper. The radial wave function can be decomposed into dipole and nondipole components as
\begin{equation}
\Phi_{\ell m}  =\Phi_{\ell m}^{(\mathrm{d})} + \Phi_{\ell m}^{(\mathrm{nd})}.
\end{equation}
Exploiting the relations for spherical harmonics  and Clebsch--Gordan coefficients, see Appendix~\ref{app_A}, the dynamics of the nondipole wave function is governed by the following equation:

\begin{equation}
\scalebox{0.9}{$
\begin{aligned}
i \partial_t \Phi_{\ell m}^{(\mathrm{nd})}= 
\frac{i}{2c} \sum_{\ell' m'} \Biggr\{ \Biggl[ &
\tilde{E}^* \, \delta_{m,m'+1} 
\Biggl(
\delta_{\ell,\ell'+2} \frac{1}{2\ell-1} 
\sqrt{ 
\frac{(\ell+m-2)(\ell+m-1)(\ell+m)(\ell-m)}
{(2\ell+1)(2\ell-3)} 
}
\\
& \quad - 
\delta_{\ell,\ell'-2} \frac{1}{2\ell+3} 
\sqrt{ 
\frac{(\ell+m+1)(\ell-m+1)(\ell-m+2)(\ell-m+3)}
{(2\ell+1)(2\ell+5)} 
}
\Biggr)
\\
& - \tilde{E}^* \, \delta_{m,m'+1} \, \delta_{\ell,\ell'}
\frac{(1-2m)}{(2\ell-1)(2\ell+3)} 
\sqrt{ (\ell+m) (\ell-m+1)}
\\
& - \tilde{E} \, \delta_{m,m'-1}
\Biggl(
\delta_{\ell,\ell'+2} \frac{1}{2\ell-1}
\sqrt{ 
\frac{(\ell-m-2)(\ell-m-1)(\ell-m)(\ell+m)}
{(2\ell+1)(2\ell-3)} 
}
\\
& \quad - 
\delta_{\ell,\ell'-2} \frac{1}{2\ell+3}
\sqrt{ 
\frac{(\ell-m+1)(\ell+m+1)(\ell+m+2)(\ell+m+3)}
{(2\ell+1)(2\ell+5)} 
}
\Biggr)
\\
& + \tilde{E} \, \delta_{m,m'-1} \, \delta_{\ell,\ell'}
\frac{(1+2m)}{(2\ell-1)(2\ell+3)} 
\sqrt{(\ell-m)(\ell+m+1)}
\Biggr]
\sqrt{r} \, \partial_r \, \sqrt{r}
\\
& +
\tilde{E}^* \, \delta_{m,m'+1}
\Biggl[
- \delta_{\ell,\ell'+2}\frac{1}{2\ell - 1}
\sqrt{
\frac{(\ell+m-1)(\ell+m)(\ell-m)}
{(2\ell+1)(2\ell-3)}
}
\Bigl(
(m + \tfrac{1}{2})\sqrt{(\ell + m - 2)}
\\
& \quad + \sqrt{(\ell - m - 1)\bigl[(\ell-2)(\ell-1) - (m -1)m \bigr]}
\Bigr)
\\
& \quad + \delta_{\ell,\ell'-2}\frac{1}{2\ell + 3}
\sqrt{
\frac{(\ell+m+1)(\ell-m+1)(\ell-m+2)}
{(2\ell+1)(2\ell+5)}
}
\Bigl(
(m + \tfrac{1}{2})\sqrt{(\ell-m+3)}
\\
& \quad - \sqrt{(\ell+m+2)\bigl[(\ell+2)(\ell+3) - (m-1)m \bigr]}
\Bigr)
\Biggr]
\\
& + \tilde{E} \, \delta_{m,m'-1}
\Biggl[
\delta_{\ell,\ell'+2}\frac{1}{2\ell - 1}
\sqrt{
\frac{(\ell-m-1)(\ell-m)(\ell+m)}
{(2\ell+1)(2\ell-3)}
}
\Bigl(
( \tfrac{1}{2}-m)\sqrt{(\ell - m - 2)}
\\
& \quad + \sqrt{(\ell + m - 1)\bigl[(\ell-2)(\ell-1) - (m+1)m \bigr]}
\Bigr)
\\
& \quad + \delta_{\ell,\ell'-2}\frac{1}{2\ell + 3}
\sqrt{
\frac{(\ell-m+1)(\ell+m+1)(\ell+m+2)}
{(2\ell+1)(2\ell+5)}
}
\Bigl(
(m - \tfrac{1}{2})\sqrt{(\ell+m+3)}
\\
& \quad + \sqrt{(\ell-m+2)\bigl[(\ell+2)(\ell+3) - (m+1)m \bigr]}
\Bigr)
\Biggr]
\\
&+ \tilde{E}^{*}\,\delta_{m,m'+1}\,\delta_{\ell,\ell'}
\Biggl[
    \frac{\left(m+\frac12\right)(1-2m)}
         {(2\ell-1)(2\ell+3)}
    \sqrt{(\ell+m)(\ell-m+1)}
\\
&\qquad\qquad
    -
    \left(
        \frac{2\bigl(\ell(\ell+1)-3m^2\bigr)}
             {3(2\ell-1)(2\ell+3)}
        + \frac13
    \right)
    \sqrt{\ell(\ell+1)-(m-1)m}
\Biggr]
\\[1ex]
&+ \tilde{E}\,\delta_{m,m'-1}\,\delta_{\ell,\ell'}
\Biggl[
    \frac{\left(m-\frac12\right)(1+2m)}
         {(2\ell-1)(2\ell+3)}
    \sqrt{(\ell-m)(\ell+m+1)}
\\
&\qquad\qquad
    +
    \left(
        \frac{2\bigl(\ell(\ell+1)-3m^2\bigr)}
             {3(2\ell-1)(2\ell+3)}
        + \frac13
    \right)
    \sqrt{\ell(\ell+1)-m(m+1)}
\Biggr]
\\
& -2ir\tilde{A}\tilde{E}^{*} \Biggl(\delta_{m,m'} \, \delta_{\ell,\ell'+1}
\sqrt{\frac{(\ell+m)(\ell-m)}{(2\ell - 1)(2\ell + 1)}}
+\delta_{m,m'} \, \delta_{\ell,\ell'-1}\sqrt{ \frac{(\ell - m +1)(\ell + m + 1)}{(2\ell+1)(2\ell+3)} }\Biggr)\Biggr\}\Phi_{\ell' m'}.
\end{aligned}
$}
\end{equation}

\section{Nondipole Corrections for Qprop}\label{sec:rel}

 \subsection{Matrix representation of the Nondipole Hamiltonian}
Equation~(11) can be written in the matrix form as
\begin{equation}
\begin{aligned}
i \partial_t \mathbf{\Phi} = H \mathbf{\Phi},
\end{aligned}
\end{equation}
with
\begin{equation}
\begin{aligned}
H = H_{\mathrm{at}}+H_{\mathrm{d}}+H_{\mathrm{nd}}^{\mathrm{mix}}+H_{\mathrm{nd}}^{\mathrm{ang}}.
\end{aligned}
\end{equation}

The matrix components of $H_{\mathrm{nd}}^{\mathrm{mix}}$ and $H_{\mathrm{nd}}^{\mathrm{ang}}$  are given by
\begin{equation}
\begin{aligned}
[H_{\mathrm{nd}}^{\mathrm{mix}}]^{\ell' m'}_{\ell m} 
&= 
\Big(
  \mathcal{A}_{\ell m}\,\delta_{m,m'+1}\delta_{\ell,\ell'+2}
+ \mathcal{B}_{\ell m}\,\delta_{m,m'+1}\delta_{\ell,\ell'-2}
+ \mathcal{C}_{\ell m}\,\delta_{m,m'+1}\delta_{\ell,\ell'} \nonumber\\
&\quad
+ \tilde{\mathcal{A}}_{\ell m}\,\delta_{m,m'-1}\delta_{\ell,\ell'+2}
+ \tilde{\mathcal{B}}_{\ell m}\,\delta_{m,m'-1}\delta_{\ell,\ell'-2}
+ \tilde{\mathcal{C}}_{\ell m}\,\delta_{m,m'-1}\delta_{\ell,\ell'}
\Big)\,
\sqrt{r}\,\partial_r\,\sqrt{r}
\\[6pt]
[H_{\mathrm{nd}}^{\mathrm{ang}}]^{\ell' m'}_{\ell m} 
&= \mathbf{1}_r \otimes 
\Big(
  \mathcal{D}_{\ell m}\,\delta_{m,m'+1}\delta_{\ell,\ell'+2}
 + \mathcal{E}_{\ell m}\,\delta_{m,m'+1}\delta_{\ell,\ell'-2} \nonumber
 + \mathcal{F}_{\ell m}\,\delta_{m,m'+1}\delta_{\ell,\ell'} 
 + \mathcal{G}_{\ell m}\,\delta_{m,m'}\delta_{\ell,\ell'+1} \nonumber\\
&\quad + \tilde{\mathcal{D}}_{\ell m}\,\delta_{m,m'-1}\delta_{\ell,\ell'+2}
 + \tilde{\mathcal{E}}_{\ell m}\,\delta_{m,m'-1}\delta_{\ell,\ell'-2}
 + \tilde{\mathcal{F}}_{\ell m}\,\delta_{m,m'-1}\delta_{\ell,\ell'} \nonumber
 + \tilde{\mathcal{G}}_{\ell m}\,\delta_{m,m'}\delta_{\ell,\ell'-1} 
\Big),
\end{aligned}
\end{equation}
where $\mathcal{A}_{\ell m}$, $\tilde{\mathcal{A}}_{\ell m}$, $\mathcal{B}_{\ell m}$, $\tilde{\mathcal{B}}_{\ell m}$, $\mathcal{C}_{\ell m}$, $\tilde{\mathcal{C}}_{\ell m}$, $\mathcal{D}_{\ell m}$, $\tilde{\mathcal{D}}_{\ell m}$, $\mathcal{E}_{\ell m}$, $\tilde{\mathcal{E}}_{\ell m}$, $\mathcal{F}_{\ell m}$, $\tilde{\mathcal{F}}_{\ell m}$, $\mathcal{G}_{\ell m}$, $\tilde{\mathcal{G}}_{\ell m}$ are the corresponding coefficients in Eq (16).  The wave function can be represented as a vector
\[
\boldsymbol{\Phi} =
\begin{pmatrix}
\fbox{$\Phi_{00}$},
\fbox{$\Phi_{1,-1}, \Phi_{10}, \Phi_{11}$},
\fbox{$\Phi_{2,-2}, \Phi_{2,-1}, \Phi_{20}, \Phi_{21}, \Phi_{22}$},
\dots,
\fbox{$\Phi_{L,L-1}, \Phi_{LL}$}
\end{pmatrix}^{\!\!\top}
\]
With the sub-blocks indicated by boxes and the discretized value of $r$ held fixed, both 
$H_{\mathrm{nd}}^{\mathrm{mix}}$ and $H_{\mathrm{nd}}^{\mathrm{ang}}$ 
can be expressed as sums of $2\times 2$ matrices acting within each $(\ell, m)$ subspace:
\begin{align}
H_{\text{nd}}^{mix} &=
\sum_{\ell=0}^{L-1} \sum_{m=-\ell}^{\ell-1}
H^{(\ell,m)}_{\text{nd,mix(1)}} 
+ \sum_{\ell=0}^{L-3} \sum_{m=-\ell}^{\ell}
\left( H^{(\ell,m)}_{\text{nd,mix(2)}} + H^{(\ell,m)}_{\text{nd,mix(3)}} \right),
\end{align}
\begin{align}
H_{\text{nd}}^{ang} &=
\sum_{\ell=0}^{L-2} \sum_{m=-\ell}^{\ell}
H^{(\ell,m)}_{\text{nd,ang(1)}} \notag \\
&\quad + \sum_{\ell=0}^{L-1} \sum_{m=-\ell}^{\ell-1}
H^{(\ell,m)}_{\text{nd,ang(2)}} \notag \\
&\quad + \sum_{\ell=0}^{L-3} \sum_{m=-\ell}^{\ell}
\left( H^{(\ell,m)}_{\text{nd,ang(3)}} + H^{(\ell,m)}_{\text{nd,ang(4)}} \right),
\end{align}
with 
\begin{equation}
H^{(\ell,m)}_{\text{nd,mix(1)}} 
= -\frac{i|\tilde{E}|}{2c}g_{\ell m}
\left(
\begin{array}{c|cc}
 & \ell m & \ell, m + 1 \\ \hline
\ell m & 0 & \exp(i\phi)\\ 
\ell, m + 1 & \exp(-i\phi) & 0
\end{array}
\right)
\otimes \sqrt{r}\partial_r\sqrt{r}
\label{glm}
\end{equation}
\begin{equation}
H^{(\ell,m)}_{\text{nd,mix(2)}} 
= \frac{i|\tilde{E}|}{2c}h_{\ell m}
\left(
\begin{array}{c|cc}
 & \ell m & \ell + 2, m + 1 \\ \hline
\ell m & 0 & \exp(i\phi)\\ 
\ell + 2, m + 1 & \exp(-i\phi) & 0
\end{array}
\right)
\otimes \sqrt{r}\partial_r\sqrt{r}
\label{hlm}
\end{equation}
\begin{equation}
H^{(\ell,m)}_{\text{nd,mix(3)}} 
= -\frac{i|\tilde{E}|}{2c}f_{\ell m}
\left(
\begin{array}{c|cc}
 & \ell m & \ell + 2, m - 1 \\ \hline
\ell m & 0 & \exp(-i\phi)\\ 
\ell + 2, m - 1 & \exp(i\phi) & 0
\end{array}
\right)
\otimes \sqrt{r}\partial_r\sqrt{r}
\label{flm}
\end{equation}
\begin{equation}
H^{(\ell,m)}_{\text{nd,ang(1)}} 
= \frac{|\tilde{A}\tilde{E}^*|}{c}p_{\ell m}
\left(
\begin{array}{c|cc}
 & \ell m & \ell+1, m \\ \hline
\ell m & 0 & 1\\ 
\ell+1, m & 1 & 0
\end{array}
\right)
\otimes r
\label{plm}
\end{equation}
\begin{equation}
H^{(\ell,m)}_{\text{nd,ang(2)}} 
= \frac{i|\tilde{E}|}{2c}n_{\ell m}
\left(
\begin{array}{c|cc}
 & \ell m & \ell, m + 1 \\ \hline
\ell m & 0 & \exp(-i\phi)\\ 
\ell, m + 1 & -\exp(i\phi) & 0
\end{array}
\right)
\label{nlm}
\end{equation}
\begin{equation}
H^{(\ell,m)}_{\text{nd,ang(3)}} 
= \frac{i|\tilde{E}|}{2c}\tilde{e}_{\ell m}
\left(
\begin{array}{c|cc}
 & \ell m & \ell + 2, m + 1 \\ \hline
\ell m & 0 & \exp(i\phi)\\ 
\ell + 2, m + 1 & -\exp(-i\phi) & 0
\end{array}
\right)
\label{e~lm}
\end{equation}
\begin{equation}
H^{(\ell,m)}_{\text{nd,ang(4)}} 
= \frac{i|\tilde{E}|}{2c}e_{\ell m}
\left(
\begin{array}{c|cc}
 & \ell m & \ell + 2, m - 1 \\ \hline
\ell m & 0 & \exp(-i\phi)\\ 
\ell + 2, m - 1 & -\exp(i\phi) & 0
\end{array}
\right)
\label{elm}.
\end{equation}

The phase $\phi$ is defined through $\tilde{E} = |\tilde{E}| \exp(i\phi)$, the coefficients ${g}_{\ell m},{h}_{\ell m},{f}_{\ell m},{p}_{\ell m},{n}_{\ell m}\tilde{e}_{\ell m},{e}_{\ell m}$ are given in   Appendix~\ref{app_B}.  Using the Numerov and Simpson approximations of the first derivative of a function $f(r)$, one obtains
\begin{equation}
\frac{\partial f(r)}{\partial r} = M_1^{-1} \Delta_1 f(r), 
\end{equation}
where \( \Delta_1 f = (f_{n+1} - f_{n-1}) / 2h \) , \( h = \Delta r \) and
\[
M_1 = \frac{1}{6}
\begin{pmatrix}
4 & 1 & 0 & \cdots \\
1 & 4 & 1 & \\
0 & 1 & \ddots & \\
\vdots & & 1 & 4
\end{pmatrix}.
\]
The corner elements of \(\Delta_1\) and \(M_1\) need to be modified, 
because the operator \(\tilde{M}_1^{-1} \tilde{\Delta}_1\) must be anti-Hermitian 
in order to ensure unitary time propagation. 
The corrected matrix forms are as follows:
\[
\tilde{M}_1 = \frac{1}{6}
\begin{pmatrix}
\sqrt{3}+2 & 1 & 0 & \cdots \\
1 & 4 & 1 & \\
0 & 1 & \ddots & \\
\vdots & & 1 & \sqrt{3}+2
\end{pmatrix},
\qquad
\tilde{\Delta}_1 = \frac{1}{2 h}
\begin{pmatrix}
\sqrt{3}-2 & 1 & 0 & \cdots \\
-1 & 0 & 1 & \\
0 & -1 & \ddots & \\
\vdots & & -1 & 2-\sqrt{3}
\end{pmatrix}.
\]
Therefore,
\begin{equation}
\sqrt{r}\frac{\partial}{\partial r}\sqrt{r}
= \sqrt{r} \tilde{M}_1^{-1} \Delta_1 \sqrt{r}
= \left( \tilde{M}_1 (\sqrt{r})^{-1} \right)^{-1}
\left( \tilde{\Delta}_1 \sqrt{r} \right)
= \tilde{M}_2^{-1} \tilde{\Delta}_2,
\end{equation}
where
\[
\tilde{M}_2 = \tilde{M}_1 (\sqrt{r})^{-1}
= \frac{1}{6}
\begin{pmatrix}
\dfrac{\sqrt{3}+2}{\sqrt{r_1}} & \dfrac{1}{\sqrt{r_2}} & 0 & \cdots \\
\dfrac{1}{\sqrt{r_1}} & \dfrac{4}{\sqrt{r_2}} & \dfrac{1}{\sqrt{r_3}} & \\
0 & \dfrac{1}{\sqrt{r_2}} & \ddots & \\
\vdots & & & \dfrac{\sqrt{3}+2}{\sqrt{r_N}}
\end{pmatrix},
\]
and
\[
\tilde{\Delta}_2 = \Delta_1 \sqrt{r}
= \frac{1}{2h}
\begin{pmatrix}
\dfrac{\sqrt{3}-2}{(\sqrt{r_1})^{-1}} & \sqrt{r_2} & 0 & \cdots \\
-\sqrt{r_1} & 0 & \sqrt{r_3} & \\
0 & -\sqrt{r_2} & \ddots & \\
\vdots & & & \dfrac{2-\sqrt{3}}{(\sqrt{r_N})^{-1}}
\end{pmatrix}.
\]

\subsection{Splitting of the nondipole time-evolution operator}\label{sec:phaseRel}

A second-order accurate approximation of the nondipole time-evolution operator can be written as
\begin{align}
U_{\text{split}}(t+\Delta t, t)
&=
\prod_{\ell=L-3}^{0} \prod_{m=\ell}^{-\ell}
\Big[
\exp\!\left(-i \tau H_{\text{nd,ang}(4)}^{\ell m}\right)
\exp\!\left(-i \tau  H_{\text{nd,ang}(3)}^{\ell m}\right)
\Big]
\notag \\[0.5em]
&\quad\times
\prod_{\ell=L-1}^{0} \prod_{m=\ell-1}^{-\ell}
\exp\!\left(-i \tau  H_{\text{nd,ang}(2)}^{\ell m}\right)
\notag \\[0.5em]
&\quad\times
\prod_{\ell=L-2}^{0} \prod_{m=\ell}^{-\ell}
\exp\!\left(-i \tau  H_{\text{nd,ang}(1)}^{\ell m}\right)
\notag \\[0.5em]
&\quad\times
\prod_{\ell=L-3}^{0} \prod_{m=\ell}^{-\ell}
\Big[
\exp\!\left(-i \tau  H_{\text{nd,mix}(3)}^{\ell m }\right)
\exp\!\left(-i \tau  H_{\text{nd,mix}(2)}^{\ell m }\right)
\Big]
\notag \\[0.5em]
&\quad\times
\prod_{\ell=L-1}^{0} \prod_{m=\ell-1}^{-\ell}
\exp\!\left(-i \tau  H_{\text{nd,mix}(1)}^{ \ell m }\right)
\notag \\[0.5em]
&\quad\times
\prod_{\ell=L-2}^{0} \prod_{m=\ell}^{-\ell}
\exp\!\left(-i \tau  H_{\text{d}}^{\ell m}\right)
\notag \\[0.5em]
&\quad\times
\exp(-i \Delta t\, H_0)
\notag \\[0.5em]
&\quad\times
\prod_{\ell=0}^{L-2} \prod_{m=-\ell}^{\ell}
\exp\!\left(-i \tau  H_{\text{d}}^{\ell m}\right)
\notag \\[0.5em]
&\quad\times
\prod_{\ell=0}^{L-1} \prod_{m=-\ell}^{\ell-1}
\exp\!\left(-i \tau  H_{\text{nd,mix}(1)}^{\ell m}\right)
\notag \\[0.5em]
&\quad\times
\prod_{\ell=0}^{L-3} \prod_{m=-\ell}^{\ell}
\Big[
\exp\!\left(-i \tau  H_{\text{nd,mix}(2)}^{ \ell m }\right)
\exp\!\left(-i \tau  H_{\text{nd,mix}(3)}^{ \ell m}\right)
\Big]
\notag \\[0.5em]
&\quad\times
\prod_{\ell=0}^{L-2} \prod_{m=-\ell}^{\ell}
\exp\!\left(-i \tau  H_{\text{nd,ang}(1)}^{ \ell m}\right)
\notag \\[0.5em]
&\quad\times
\prod_{\ell=0}^{L-1} \prod_{m=-\ell}^{\ell-1}
\exp\!\left(-i \tau  H_{\text{nd,ang}(2)}^{ \ell m}\right)
\notag \\[0.5em]
&\quad\times
\prod_{\ell=0}^{L-3} \prod_{m=-\ell}^{\ell}
\Big[
\exp\!\left(-i \tau  H_{\text{nd,ang}(3)}^{ \ell m }\right)
\exp\!\left(-i \tau  H_{\text{nd,ang}(4)}^{ \ell m}\right)
\Big].
\end{align}
With $\tau = \Delta t / 2$, Equation (31) can be further discretized using the Crank-Nicolson approximant
\begin{align}
U_{\mathrm{split}}(t + \Delta t, t) &\approx \mathbf{U}_{\mathrm{CN}}(t + \Delta t, t) \notag \\ 
&= \prod_{\ell = L - 3}^{0} \prod_{m=\ell}^{-\ell} \left(\mathbf{E}_n^{\ell m} \mathbf{X}_-^{\ell m} \left[ \mathbf{X}_+^{\ell m} \right]^{-1} \mathbf{\tilde{E}}_n^{\ell m} \mathbf{\tilde{X}}_-^{\ell m} \left[ \mathbf{\tilde{X}}_+^{\ell m} \right]^{-1}\right)
   \prod_{\ell = L - 1}^{0} \prod_{m=\ell-1}^{-\ell} \left( \mathbf{N}_n^{\ell m} \mathbf{L}_-^{\ell m} \left[ \mathbf{L}_+^{\ell m} \right]^{-1}\right) \notag \\
&\quad \times \prod_{\ell=L-2}^{0} \prod_{m=\ell}^{-\ell} \left(H_n^{\ell m} D_n^{\ell m} \right) \, \mathbf{Q}_+^{-1} \, \mathbf{Q}_- 
   \prod_{\ell=0}^{L-2} \prod_{m=-\ell}^{\ell}  \left( D_n^{\ell m} H_n^{\ell m} \right) \notag \\
&\quad \times \prod_{\ell = 0}^{L - 1} \prod_{m=-\ell}^{\ell-1} \left( \left[ \mathbf{L}_+^{\ell m} \right]^{-1} \mathbf{L}_-^{\ell m} \, \mathbf{N}_n^{\ell m} \right) \prod_{\ell = 0}^{L - 3} \prod_{m=-\ell}^{\ell} \left( \left[ \mathbf{\tilde{X}}_+^{\ell m} \right]^{-1} \mathbf{\tilde{X}}_-^{\ell m} \, \mathbf{\tilde{E}}_n^{\ell m} \left[ \mathbf{X}_+^{\ell m} \right]^{-1} \mathbf{X}_-^{\ell m} \, \mathbf{E}_n^{\ell m} \right)
\end{align}
where
\begin{align}
\mathbf{E}_n^{\ell m} 
    &= \left( \mathbf{1} + \mathrm{i} \frac{\tau}{2} 
        H_{\mathrm{nd,ang}(4)}^{\ell m} \right)^{-1}
       \left( \mathbf{1} - \mathrm{i} \frac{\tau}{2} 
        H_{\mathrm{nd,ang}(4)}^{\ell m} \right),
    \notag \\[6pt]
\mathbf{\tilde{E}}_n^{\ell m} 
    &= \left( \mathbf{1} + \mathrm{i} \frac{\tau}{2} 
        H_{\mathrm{nd,ang}(3)}^{\ell m} \right)^{-1}
       \left( \mathbf{1} - \mathrm{i} \frac{\tau}{2} 
        H_{\mathrm{nd,ang}(3)}^{\ell m} \right),
    \notag \\[6pt]
\mathbf{N}_n^{\ell m} 
    &= \left( \mathbf{1} + \mathrm{i} \frac{\tau}{2} 
        H_{\mathrm{nd,ang}(2)}^{\ell m} \right)^{-1}
       \left( \mathbf{1} - \mathrm{i} \frac{\tau}{2} 
        H_{\mathrm{nd,ang}(2)}^{\ell m} \right),
    \notag \\[6pt]
\mathbf{H}_n^{\ell m} 
    &= \left( \mathbf{1} + \mathrm{i} \frac{\tau}{2} 
        H_{\mathrm{nd,ang}(1)}^{\ell m} \right)^{-1}
       \left( \mathbf{1} - \mathrm{i} \frac{\tau}{2} 
        H_{\mathrm{nd,ang}(1)}^{\ell m} \right),
    \notag \\[6pt]
\mathbf{D}_n^{\ell m} 
    &= \left( \mathbf{1} + \mathrm{i} \frac{\tau}{2} 
        \mathbf{H}_{\mathrm{d}}^{\ell m} \right)^{-1}
       \left( \mathbf{1} - \mathrm{i} \frac{\tau}{2} 
        \mathbf{H}_{\mathrm{d}}^{\ell m} \right),
    \notag \\[6pt]
\mathbf{X}_{\pm}^{\ell m} 
    &= \mathbf{1} \pm \mathrm{i} \frac{\tau}{2} 
       \mathbf{H}_{\mathrm{nd},\mathrm{mix}(3)}^{\ell m},
    \notag \\[6pt]
\mathbf{\tilde{X}}_{\pm}^{\ell m}
    &= \mathbf{1} \pm \mathrm{i} \frac{\tau}{2} 
       \mathbf{H}_{\mathrm{nd},\mathrm{mix}(2)}^{\ell m},
    \notag \\[6pt]
\mathbf{L}_{\pm}^{\ell m} 
    &= \mathbf{1} \pm \mathrm{i} \frac{\tau}{2} 
       \mathbf{H}_{\mathrm{nd},\mathrm{mix}(1)}^{\ell m},
    \notag \\[6pt]
\mathbf{Q}_{\pm} 
    &= \mathbf{1} \pm \mathrm{i} \tau\, \mathbf{H}_{\mathrm{at}}.
    \notag
\end{align}

$\mathbf{E}_n^{\ell m}, \mathbf{\tilde{E}}_n^{\ell m},\mathbf{N}_n^{\ell m},\mathbf{P}_n^{\ell m}$ can be evaluated as
\begin{align}
\mathbf{E}_n^{\ell m}
&= \frac{1}{1 + \xi^{2} e_{\ell m}^{2}}
\begin{pmatrix}
1 - \xi^{2} e_{\ell m}^{2} & 2\xi\, e^{-i\phi} e_{\ell m} \\
-2\xi\, e^{i\phi} e_{\ell m} & 1 - \xi^{2} e_{\ell m}^{2}
\end{pmatrix},
\notag\\[6pt]
\mathbf{\tilde{E}}_n^{\ell m}
&= \frac{1}{1 + \xi^{2} \tilde{e}_{\ell m}^{2}}
\begin{pmatrix}
1 - \xi^{2} \tilde{e}_{\ell m}^{2} & 2\xi\, e^{i\phi} \tilde{e}_{\ell m} \\
-2\xi\, e^{-i\phi} \tilde{e}_{\ell m} & 1 - \xi^{2} \tilde{e}_{\ell m}^{2}
\end{pmatrix},
\notag\\[6pt]
\mathbf{N}_n^{\ell m}
&= \frac{1}{1 + \xi^{2} n_{\ell m}^{2}}
\begin{pmatrix}
1 - \xi^{2} n_{\ell m}^{2} & 2\xi\, e^{-i\phi} n_{\ell m} \\
-2\xi\, e^{i\phi} n_{\ell m} & 1 - \xi^{2} n_{\ell m}^{2}
\end{pmatrix},
\notag\\[6pt]
\mathbf{P}_n^{\ell m}
&= \frac{1}{1 + \zeta^{2} p_{\ell m}^{2}}
\begin{pmatrix}
1 - \zeta^{2} p_{\ell m}^{2} & 2 \zeta p_{\ell m} \\
-2\zeta\, p_{\ell m} & 1 - \zeta^{2} p_{\ell m}^{2}
\end{pmatrix}.
\notag\
\end{align}
With $\xi = \tau |\tilde{E}|/(4c)$ and $\zeta = \tau |\tilde{A}\tilde{E}^{*}|\, r/(2c)$, we now turn to the factor $\left[ \mathbf{X}_+^{\ell m} \right]^{-1}\mathbf{X}_-^{\ell m}$, which couples the $(\ell,\ell+2)$ sub-blocks and introduces a non-diagonal structure in the radial coordinate $r$. The treatment of $\left[ \mathbf{\tilde{L}}_+^{\ell m} \right]^{-1}\mathbf{\tilde{L}}_-^{\ell m}$ proceeds analogously, with the substitution of the coefficient $f_{\ell m}$ by $g_{\ell m}$ and by working with the $(m,m+1)$ sub-blocks instead. To facilitate the inversion, we factor out $\tilde{\mathbf{M}}_{2}^{-1}$ and rewrite the expression accordingly:
\[
\left[ \mathbf{X}_+^{\ell m} \right]^{-1} \mathbf{X}_-^{\ell m} = \left[ \mathbf{Y}_+^{\ell m} \right]^{-1} \mathbf{Y}_-^{\ell m}, \quad \mathbf{Y}_{\pm}^{\ell m} = \mathbf{1}_{\ell} \otimes \tilde{\mathbf{M}}_2 \pm \frac{\tau}{4c} |\tilde{E}| f_{\ell m} P \otimes \tilde{\boldsymbol{\Delta}}_2 
\]

\[
\left[ \mathbf{\tilde{X}}_+^{\ell m} \right]^{-1} \mathbf{\tilde{X}}_-^{\ell m} = \left[ \mathbf{\tilde{Y}}_+^{\ell m} \right]^{-1} \mathbf{\tilde{Y}}_-^{\ell m}, \quad \mathbf{\tilde{Y}}_{\pm}^{\ell m} = \mathbf{1}_{\ell} \otimes \tilde{\mathbf{M}}_2 \mp \frac{\tau}{4c} |\tilde{E}| h_{\ell m} \tilde{P} \otimes \tilde{\boldsymbol{\Delta}}_2 ,
\]
where
\begin{equation}
P = 
\left(
\begin{array}{c c}
 0 & \exp(-i\phi)\\ 
 \exp(i\phi) & 0
 \notag\
\end{array}
\right)
\end{equation}
\begin{equation}
\tilde{P} = 
\left(
\begin{array}{c c}
 0 & \exp(i\phi)\\ 
 \exp(-i\phi) & 0
 \notag\
\end{array}
\right).
\end{equation}
Since
\[
\mathbf{B}
= \frac{1}{\sqrt{2}}
\begin{pmatrix}
-\,\exp(i\phi) & 1 \\
\exp(i\phi) & 1
\end{pmatrix},
\qquad
\mathbf{B}^{-1} = \mathbf{B}^{\dagger}
= \frac{1}{\sqrt{2}}
\begin{pmatrix}
-\,\exp(-i\phi) & \exp(-i\phi) \\
1 & 1
\end{pmatrix}.
\]

\[
\tilde{\mathbf{B}}
= \frac{1}{\sqrt{2}}
\begin{pmatrix}
-\,\exp(-i\phi) & 1 \\
\exp(-i\phi) & 1
\end{pmatrix},
\qquad
\tilde{\mathbf{B}}^{-1} = \tilde{\mathbf{B}}^{\dagger}
= \frac{1}{\sqrt{2}}
\begin{pmatrix}
-\,\exp(i\phi) & \exp(i\phi) \\
1 & 1
\end{pmatrix}.
\]
one obtains
\[ 
\text{BPB}^\dagger = {\tilde{B}\tilde{P}\tilde{B}}^\dagger = \begin{pmatrix} -1 & 0 \\ 0 & 1 \end{pmatrix}.
\]
It follows that
\[
\left[ \mathbf{Y}_+^{\ell m} \right]^{-1} \mathbf{Y}_-^{\ell m}
= \text{B}^\dagger \text{B} \left[ Y_{+}^{\ell m} \right]^{-1} \text{B}^\dagger \text{B} Y_{-}^{\ell m} \text{B}^\dagger \text{B}
\]

\[
= \text{B}^\dagger \left[ \text{B} Y_{+}^{\ell m} \text{B}^\dagger \right]^{-1} \text{B} Y_{-}^{\ell m} \text{B}^\dagger \text{B}
\]

\[
= \text{B}^\dagger\underbrace{ \left[\hat{M}_2 + \zeta f_{\ell m} \begin{pmatrix} -1 & 0 \\ 0 & 1 \end{pmatrix} \tilde{\Delta}_2\right]}_{\overline{Y}_{+}^{\ell m}} \underbrace{\left[ \hat{M}_2 - \zeta f_{\ell m} \begin{pmatrix} -1 & 0 \\ 0 & 1 \end{pmatrix} \tilde{\Delta}_2 \right]}_{\overline{Y}_{-}^{\ell m}} \text{B},
\]

\[
\left[ \mathbf{\tilde{Y}}_+^{\ell m} \right]^{-1} \mathbf{\tilde{Y}}_-^{\ell m}
= \tilde{B}^\dagger\underbrace{ \left[\hat{M}_2 - \zeta f_{\ell m} \begin{pmatrix} -1 & 0 \\ 0 & 1 \end{pmatrix} \tilde{\Delta}_2\right]}_{\overline{\tilde{Y}}_{+}^{\ell m}} \underbrace{\left[ \hat{M}_2 + \zeta f_{\ell m} \begin{pmatrix} -1 & 0 \\ 0 & 1 \end{pmatrix} \tilde{\Delta}_2 \right]}_{\overline{\tilde{Y}}_{-}^{\ell m}} \tilde{B}.
\]
The matrices $\mathbf{\overline{Y}}_{\pm}^{\ell m}$ are not tridiagonal but block-tridiagonal only:
\[
\mathbf{\overline{Y}}_{\pm}^{\ell m} =
\left(
\begin{array}{cc|cc|cc|cc}
\dfrac{4+x}{6\sqrt{r_1}} & \pm y g_{\ell m}\sqrt{r_1} & 
\dfrac{1}{6\sqrt{r_2}} & \pm g_{\ell m} \sqrt{r_2} & 
 & & 
 & \\[10pt]
\pm y g_{\ell m}\sqrt{r_1} & \dfrac{4+x}{6\sqrt{r_1}} & 
\pm g_{\ell m}\sqrt{r_2} & \dfrac{1}{6\sqrt{r_2}} & 
 & & 
 & \\[10pt]
\hline
\dfrac{1}{6\sqrt{r_1}} & \mp g_{\ell m}\sqrt{r_1} & 
\dfrac{2}{3\sqrt{r_2}} &  & 
 & & 
 & \\[10pt]
\mp g_{\ell m}\sqrt{r_1} & \dfrac{1}{6\sqrt{r_1}} & 
 & \dfrac{2}{3\sqrt{r_2}} & 
 & & 
 & \\[10pt]
\hline
 & & 
 & & 
\cdot &  & 
\dfrac{1}{6\sqrt{r_N}} & \pm g_{\ell m}\sqrt{r_N} \\[10pt]
 & & 
 & & 
 & \cdot & 
\pm g_{\ell m} \sqrt{r_N} & \dfrac{1}{6\sqrt{r_N}} \\[10pt]
\hline
 & & 
 & & 
\dfrac{1}{6\sqrt{r_{N-1}}} & \mp g_{\ell m} \sqrt{r_{N-1}} & 
\dfrac{4+x}{6\sqrt{r_N}} & \mp y g_{\ell m}\sqrt{r_N} \\[10pt]
 & & 
 & & 
\mp g_{\ell m}\sqrt{r_{N-1}} & \dfrac{1}{6\sqrt{r_{N-1}}} & 
\mp y g_{\ell m}\sqrt{r_N} & \dfrac{4+x}{6\sqrt{r_N}}
\end{array}
\right)
\]
with $g_{\ell m}=\tau |\tilde{E}|f_{\ell m}/4h$, $x = y =\sqrt{3}-2$. The rank of these matrices is $2N_{r}$. It operates on vectors of the form
\[
\boldsymbol{\Gamma}_{\ell m}=\left(\boxed{\Phi_{\ell m}(r_{1},t),\Phi_{\ell+2,m}(r_{1},t)}\big{|}\boxed{\Phi_{\ell m}(r_{2},t),\Phi_{\ell+2,m}(r_{2},t)}\big{|}\cdots\right)^{\top}.
\]

The matrices $\mathbf{Y}_{\pm}^{\ell m}$ can be transformed into a sum of two tridiagonal matrices $\overline{\mathbf{Y}}_{1\pm}^{\ell m}$ and $\overline{\mathbf{Y}}_{2\pm}^{\ell m}$, acting in two distinct vector spaces. 
\[
\overline{\mathbf{Y}}_{\pm}^{\ell m}=\overline{\mathbf{Y}}_{1\pm}^{\ell m}+\overline{\mathbf{Y}}_{2\pm}^{\ell m}
\]
with
\[
\overline{\mathbf{Y}}_{1\pm}^{\ell m} = 
\left(
\begin{array}{cc|cc|cc|cc}
\dfrac{4+x}{6\sqrt{r_1}}\!\pm\! y g_{\ell m}\sqrt{r_1} & 0 & 
\dfrac{1}{6\sqrt{r_2}}\!\pm\! g_{\ell m}\sqrt{r_2} & 0 & 
 & & 
 & \\[8pt]
0 & 0 & 
0 & 0 & 
 & & 
 & \\[8pt]
\hline
\dfrac{1}{6\sqrt{r_1}}\!\mp\! g_{\ell m}\sqrt{r_1} & 0 & 
\dfrac{2}{3\sqrt{r_2}} & 0 & 
 & & 
 & \\[8pt]
0 & 0 & 
0 & 0 & 
 & & 
 & \\[8pt]
\hline
 & & 
 & & 
\cdot &  & 
\dfrac{1}{6\sqrt{r_N}}\!\mp\! g_{\ell m} \sqrt{r_N} & 0 \\[8pt]
 & & 
 & & 
 & \cdot & 
0 & 0 \\[8pt]
\hline
 & & 
 & & 
\dfrac{1}{6\sqrt{r_{N-1}}}\!\mp\! g_{\ell m} \sqrt{r_{N-1}} & 0 & 
\dfrac{4+x}{6\sqrt{r_N}}\!\mp\! y g_{\ell m}\sqrt{r_N} & 0 \\[8pt]
 & & 
 & & 
0 & 0 & 
0 & 0
\end{array}
\right)
\]

\[
\overline{\mathbf{Y}}_{2\pm}^{\ell m} = 
\left(
\begin{array}{cc|cc|cc|cc}
0 & 0 & 
0 & 0 & 
 & & 
 & \\[8pt]
0 & \dfrac{4+x}{6\sqrt{r_1}}\!\mp\! y g_{\ell m}\sqrt{r_1} & 
0 & \dfrac{1}{6\sqrt{r_2}}\!\mp\! g_{\ell m}\sqrt{r_2} & 
 & & 
 & \\[8pt]
\hline
0 & 0 & 
0 & 0 & 
 & & 
 & \\[8pt]
0 & \dfrac{1}{6\sqrt{r_1}}\!\pm\! g_{\ell m}\sqrt{r_1} & 
0 & \dfrac{2}{3\sqrt{r_2}} & 
 & & 
 & \\[8pt]
\hline
 & & 
 & & 
\cdot &  & 
0 & 0 \\[8pt]
 & & 
 & & 
 & \cdot & 
0 & \dfrac{1}{6\sqrt{r_N}}\!\pm\! g_{\ell m}\sqrt{r_N} \\[8pt]
\hline
 & & 
 & & 
0 & 0 & 
0 & 0 \\[8pt]
 & & 
 & & 
0 & \dfrac{1}{6\sqrt{r_{N-1}}}\!\pm\! g_{\ell m}\sqrt{r_{N-1}} & 
0 & \dfrac{4+x}{6\sqrt{r_N}}\!\pm\! y g_{\ell m}\sqrt{r_N}
\end{array}
\right)
\]

The vector $ \Gamma_{\ell m}$ can be transformed accordingly,
\[
\overline{\Gamma}_{\ell m}=(\mathsf{B}\otimes 1_{r})\Gamma_{\ell m}=\overline{\Gamma}_{1\ell m}+\overline{\Gamma}_{2\ell m},
\]
where
\[
\overline{\Gamma}_{1\ell m}=\frac{1}{\sqrt{2}}\left(\begin{array}[]{c} 
\Phi_{\ell m}^{1}+\Phi_{\ell+2,m}^{1}\\ 
0\\ 
\hline
\Phi_{\ell m}^{2}+\Phi_{\ell+2,m}^{2}\\ 
0\\ 
\hline
\vdots\\ 
\hline
\Phi_{\ell m}^{N_{r}}+\Phi_{\ell+2,m}^{N_{r}}\\ 
0
\end{array}\right),\quad
\overline{\Gamma}_{2\ell m}=\frac{1}{\sqrt{2}}\left(\begin{array}[]{c}
0\\ 
-\Phi_{\ell m}^{1}+\Phi_{\ell+2,m}^{1}\\ 
0\\ 
\hline
-\Phi_{\ell m}^{2}+\Phi_{\ell+2,m}^{2}\\ 
\hline
\vdots\\ 
0\\ 
\hline
-\Phi_{\ell m}^{N_{r}}+\Phi_{\ell+2,m}^{N_{r}}
\end{array}\right).
\]
Now it is easy to see that
\[
\overline{Y}_{\pm}^{\ell m}\overline{\Gamma}_{\ell m}=(\overline{Y}_{1\pm}^{\ell m}+\overline{Y}_{2\pm}^{\ell m})(\overline{\Gamma}_{1\ell m}+\overline{\Gamma}_{2\ell m})=\overline{Y}_{1\pm}^{\ell m}\overline{\Gamma}_{1\ell m}+\overline{Y}_{2\pm}^{\ell m}\overline{\Gamma}_{2\ell m},
\]
because the matrices $\overline{Y}_{1\pm}^{\ell m},\overline{Y}_{2\pm}^{\ell m}$ operate in distinct spaces.

\section{Demonstration of the performance of Qprop-ND }\label{sec:examples}
\subsection{Dipole vs nondipole effects in linearly polarized laser pulses}

To  validate the accuracy of Qprop-ND, we  benchmark the results  calculated by Qprop-ND with the   relativistic calculations of Telnov and Chu~\cite{Telnov_Chu_2021}, related to the ionization in a ultrastrong XUV field in the nondipole regime. To this end, we consider the interaction of an hydrogenlike helium ion (He$^{+}$) with a linearly polarized strong laser pulse propagating in $z$ direction and polarized along the $x$ direction, with a Gaussian shape vector potential given by
\begin{align}
A_x(t) &= \frac{E_0}{\omega}
\exp\!\left[
    -2 \ln 2 \left( \frac{t}{\tau} \right)^{2}
\right] \nonumber\sin( \omega t ),
\end{align}
where
$\tau$ is the full width at half maximum (FWHM) in the intensity profile. We choose parameters similar to those of Ref.~\cite{Telnov_Chu_2021}. The laser frequency is  $\omega=14$~a.u. and the laser pulse duration  $\tau=2.5 $ optical cycles. Three cases with the peak electric field amplitude  of
$E_0=160$~a.u., $ 320$~a.u., $ 480$~a.u., are considered. The considered three cases correspond to the typical nondipole regime with the dimensionless field  parameter $a_0 \approx 0.08$, $a_0 \approx 0.17$, $a_0 \approx 0.25$.  As $a_0\ll 1$, the application of TDSE is relevant, neglecting the relativistic mass correction $\sim 1/c^2$, however, $a_0$ is not negligible such that the laser magnetic field effects cannot be neglected, at least for the two higher values of $a_0$.

\begin{figure}
    \centering
    \includegraphics[width=1\linewidth]{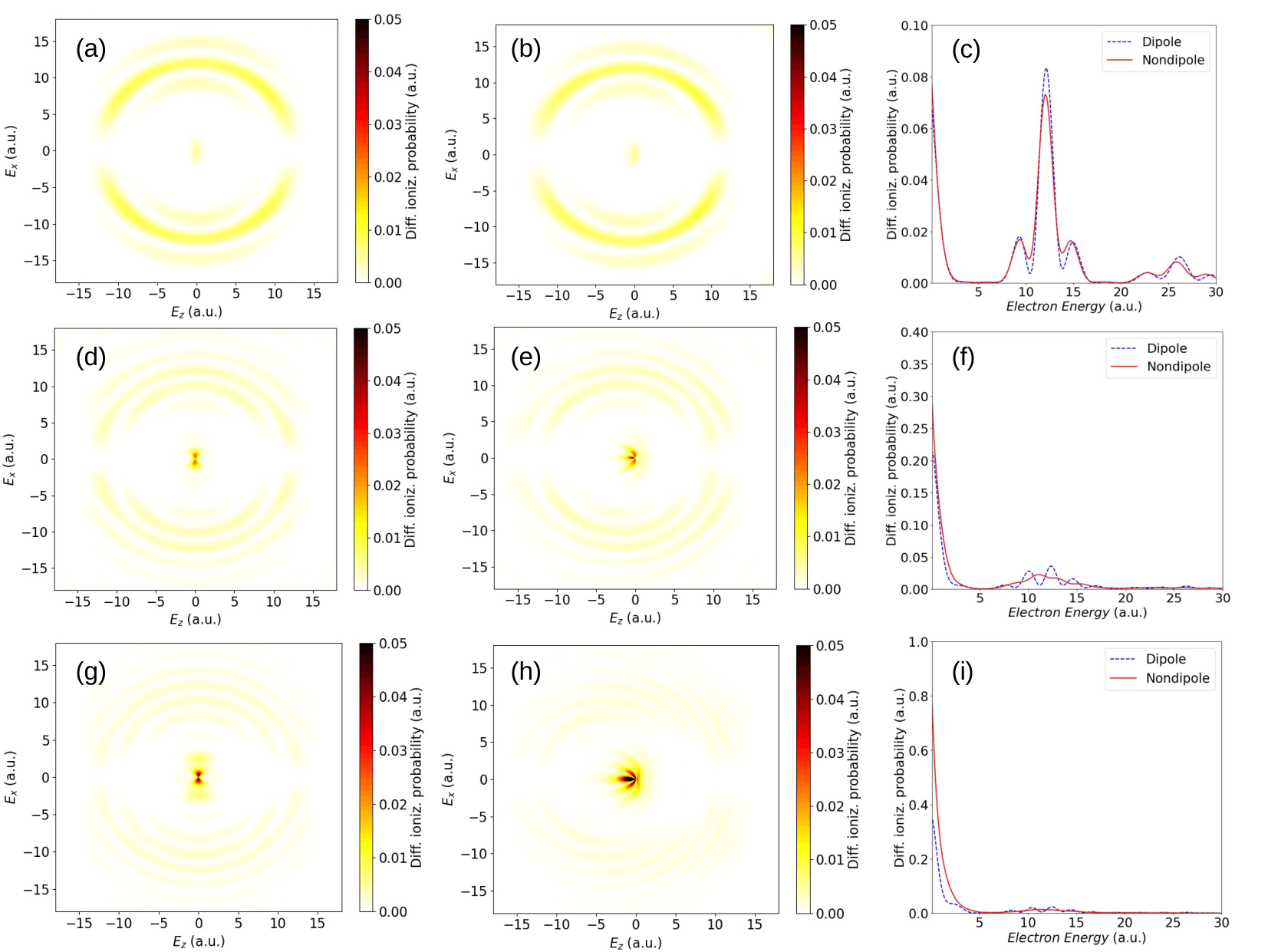}
    \caption{Comparative analysis of dipole vs nondipole approximations: The peak electric field amplitude $E_0$ of the first, second and third row in Fig 1 are: $E_0=160$~a.u., $E_0=320$~a.u., $E_0=480$~a.u., respectively. The laser frequency is set as  $\omega=14$~a.u.. The PMD calculated in the dipole approximation with Qprop 3.0 is presented in the first column, while the nondipole calculation via Qprop-ND developed in this work is shown in the second column, the relative scale of the ATI peak between dipole and nondipole is demonstrated in the third column.}
    \label{fig1}
\end{figure}

\begin{figure}
    \centering
    \includegraphics[width=1\linewidth]{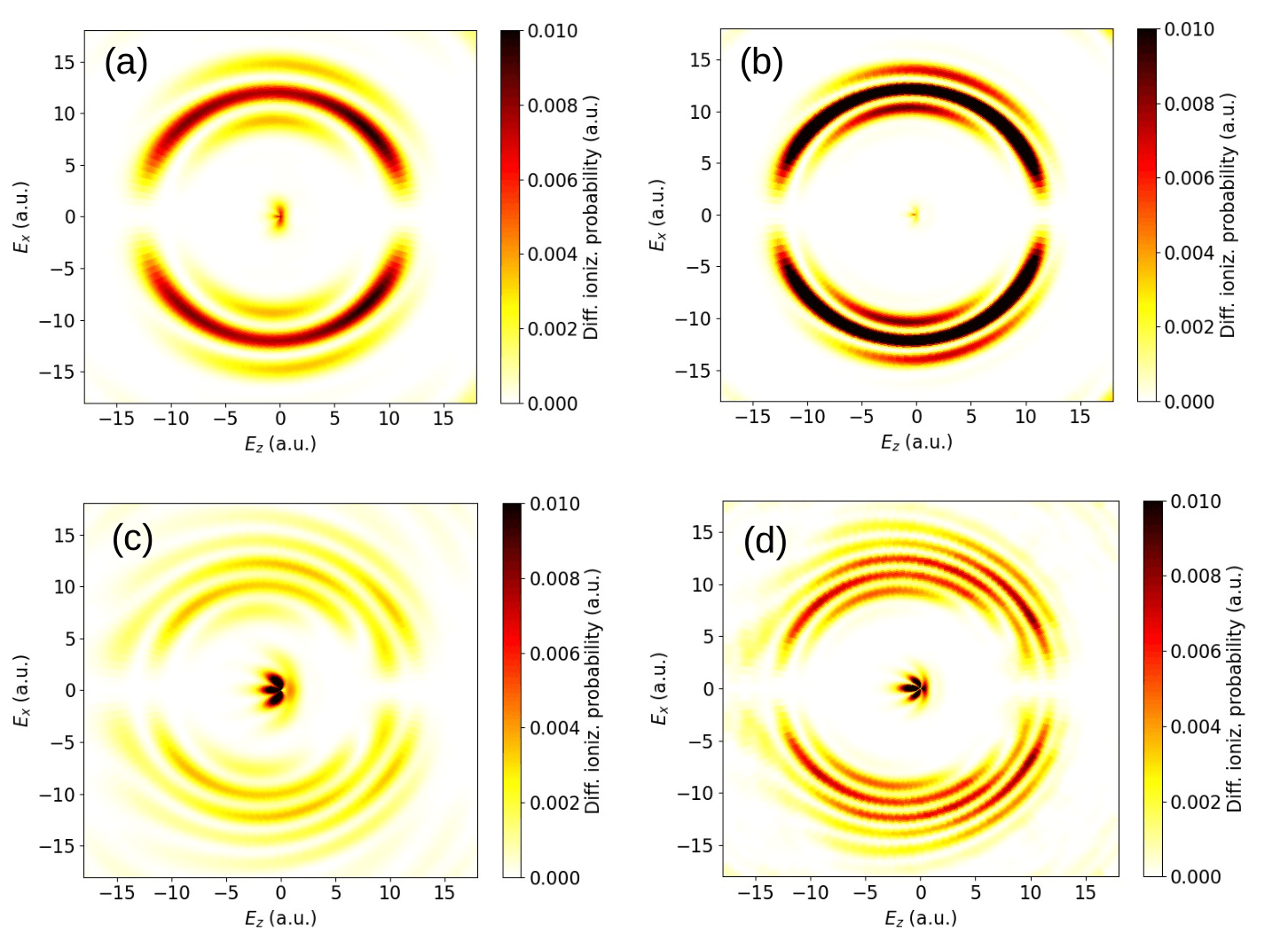}
    \caption{PMDs for different pulse durations. The peak electric field amplitudes for the upper and lower rows in Fig.~2 are $E_0=160$~a.u. and $E_0=320$~a.u., respectively. The laser frequency is fixed at $\omega=14$~a.u. The left column corresponds to a pulse duration of $\tau=2.5$ optical cycles, while the right column corresponds to $\tau=3.75$ optical cycles.}
    \label{fig2}
\end{figure}

The PMDs are presented in Fig.~\ref{fig1}. The PMDs calculated in the dipole approximation with Qprop 3.0 are shown in the first column, while the nondipole calculations via Qprop-ND developed in this work in the second column, the relative scale of the ATI peak between dipole and nondipole is demonstrated in the third column. In the dipole approximation the PMD shows low-energy two lobes along the laser electric field. In the nondipole treatment, one sees a three lobe structure at low energies, with the main lobe exactly along the laser counterpropagating direction, and two dipole lobes at a slant in the same direction. These structures are known from the literature \cite{Telnov_Chu_2021}, and can serve as evidence for the correctness for our nondipole code. For both cases the ATI rings are visible with a low probability. The ATI peaks are modulated due to dynamic interference \cite{Toyota_2008,Demekhin_2012,Jiang_2018,Geng_2021}. At a weak electric field of $E_0=160$, the PMDs and the relative ATI peak structure are similar in the dipole and nondipole spectra [cf. Fig.~\ref{fig1}(a-c) with Fig.~2 and  5 (left column) of Ref.~\cite{Telnov_Chu_2021}].  At higher field strengths the dynamic interference pattern in the nondipole ATI peaks is largely suppressed, leading to smoother lines compared with those obtained within the dipole approximation [cf. Fig.~\ref{fig1}(d-f) with Fig.~3 and  6 (left column) of Ref.~\cite{Telnov_Chu_2021}; as well as Fig.~\ref{fig1}(g-i) with Fig.~4 and  7 (left column) of Ref.~\cite{Telnov_Chu_2021}]. The centers of the first ATI rings are located near 12 a.u., which is consistent with the reported results. The zero-energy structure is enhanced with increasing the field. The ATI rings at large fields become nonuniform demonstrating asymmetry in the laser propagation direction [see zooming of Fig.~\ref{fig1}(b,e) in Fig.~\ref{fig2}(a,c)].

We further investigate the dependence of the PMDs on the pulse duration in Fig.~\ref{fig2}, following the comparison performed in Ref.~\cite{Telnov_Chu_2021}. The upper and lower rows correspond to $E_0=160$~a.u. ($a_0=0.097$) and $E_0=320$~a.u. ($a_0=0.17$), respectively, while the laser frequency is fixed at $\omega=14$~a.u. The left and right columns correspond to pulse durations of $\tau=2.5$ and $3.75$ optical cycles, respectively. The overall nondipole low-energy structure remains qualitatively unchanged as the pulse duration is increased. In contrast, the ATI distributions in Fig.~\ref{fig2}(c,d) exhibit a pronounced angular inhomogeneity, as also observed in Ref.~\cite{Telnov_Chu_2021}. Overall, the agreement in the nondipole low-energy structure and the angular inhomogeneity of the ATI distributions demonstrates that Qprop-ND reproduces the characteristic features of strong-field ionization beyond the dipole approximation.

To further validate the accuracy of Qprop-ND, we provide a quantitative comparison of our calculated ionization probabilities with the benchmark relativistic calculations of  Ref.~\cite{Telnov_Chu_2021}, see Table~\ref{tab:benchmark}. Overall, the agreement between the present calculations and the published results is very good for both the dipole and nondipole cases over a broad range of laser intensities.
For the two lower peak electric field amplitudes ($E_0=160$ and $320$~a.u.), the ionization probabilities obtained by Qprop-ND differ from the benchmark values only by a few percent. These results demonstrate that the leading-order nondipole interaction implemented in the present TDSE solver accurately reproduces the established relativistic benchmark.

A more noticeable deviation appears at the highest field strength ($E_0=480$~a.u.), where the present calculations predict systematically larger ionization probabilities and zero-energy yields, especially for the nondipole case. This difference can be understood from the different theoretical frameworks employed. Reference~\cite{Telnov_Chu_2021} solves the fully relativistic time-dependent Dirac equation (TDDE), whereas the present work is based on the nonrelativistic time-dependent Schrödinger equation including only the leading-order nondipole ($1/c$) corrections. Although the normalized vector potential remains moderate ($a_0\approx0.25$), higher-order relativistic effects, including relativistic kinematics, spin dynamics, and higher-order magnetic interactions naturally contained in the Dirac equation, become increasingly important at such strong field strengths. These effects are absent in the present TDSE formulation and therefore lead to a moderate overestimation of the ionization probability. Nevertheless, the overall agreement remains satisfactory, confirming that Qprop-ND correctly captures the dominant nondipole physics in the weakly relativistic regime and provides a reliable computational framework for the following investigations.

\begin{table}[htbp]
\caption{Comparison of the total ionization probabilities ($P_{\rm ion}$)
between the benchmark results of Telnov and Chu~\cite{Telnov_Chu_2021}
and the present Qprop and Qprop-ND calculations.}
\label{tab:benchmark}
\centering
\renewcommand{\arraystretch}{1.2}
\begin{tabular}{lccc}
\hline
Electric-field amplitude
& $E_0=160$ ($a_0=0.097$)
& $E_0=320$ ($a_0=0.17$)
& $E_0=480$ ($a_0=0.25$) \\
\hline

Ref.~\cite{Telnov_Chu_2021} Dipole
& 0.27 & 0.33 & 0.36 \\

Qprop
& 0.27 & 0.33 & 0.40 \\

Ref.~\cite{Telnov_Chu_2021} Nondipole
& 0.27 & 0.35 & 0.56 \\

Qprop-ND
& 0.27 & 0.36 & 0.57 \\

\hline
\end{tabular}
\end{table}

\subsection{Nondipole effects in the x-ray regime}

To further assess the accuracy of the present Qprop-ND implementation in the high frequency regime, we compare our calculations with the nonperturbative benchmark calculations reported by Dondera and Bachau~\cite{Dondera_2012}, where the photoionization of atomic hydrogen was investigated for photon energies ranging from 200~eV to 3~keV, by solving TDSE 
including the leading-order nondipole corrections up to $O(1/c)$. 
Although the numerical implementation differ from the present work, both approaches are based on the same weakly relativistic framework, making a direct comparison meaningful. In particular, Ref.~\cite{Dondera_2012} concluded that the nondipole effects have only a weak influence on the photoelectron energy spectra but produce much stronger modifications in the angular distributions.

\begin{figure}
    \centering
    \includegraphics[width=1\linewidth]{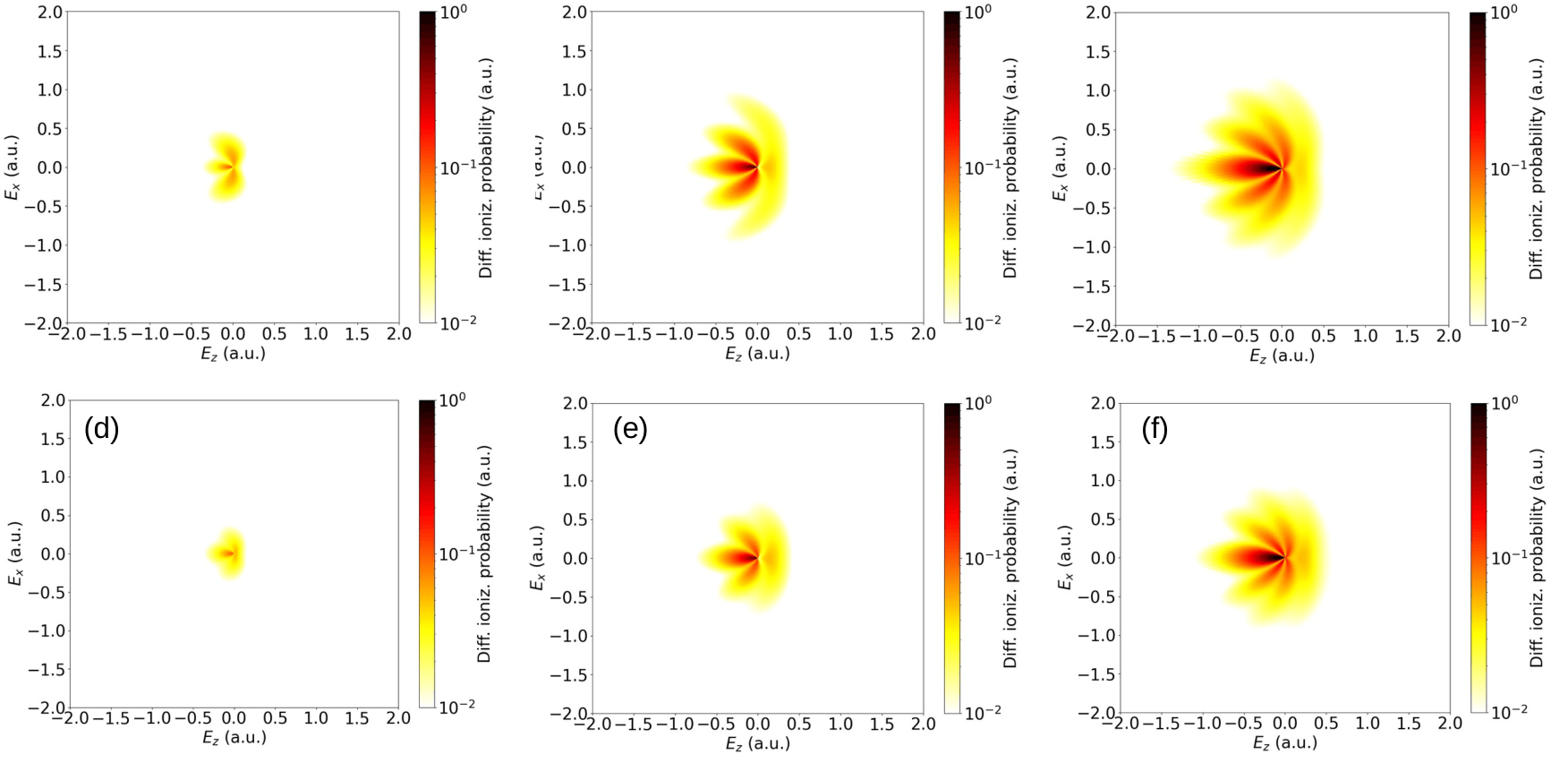}
    \caption{Comparison of the present Qprop-ND calculations with the benchmark results of Ref.~\cite{Dondera_2012}. (a) Normalized photoelectron angular distributions calculated within the dipole approximation (dashed curve) and including the leading-order nondipole corrections (solid curve). (b) Photoelectron energy spectra for photon energies of 200~eV, 500~eV, and 1~keV. The nondipole corrections become increasingly pronounced with increasing photon energy, in agreement with the nonperturbative TDSE calculations reported in Ref.~\cite{Dondera_2012}.}
    \label{Fig 3}
\end{figure}

Figure~\ref{Fig 3}(a) presents the normalized angular distributions calculated within the dipole approximation and with nondipole corrections for an x-ray pulse of peak intensity $I= 3.51 \times 10^{16}~\mathrm{W\,cm^{-2}}$ and a  photon energy of $500~\mathrm{eV}$. The pulse envelope follows a $\cos^2$ profile with a total pulse duration of 20 optical cycles of the field. The dipole calculation exhibits the characteristic $\cos^2\theta$ distribution expected for linearly polarized single-photon ionization. After including the nondipole interaction, the overall angular profile is only slightly modified. The emission probability along the polarization direction remains dominant, whereas a small redistribution of the photoelectron yield occurs at intermediate emission angles. No significant change of the angular pattern is observed, just slight asymmetry in the propagation direction, indicating that the nondipole interaction acts mainly as a perturbative correction in the present weakly relativistic regime. This behavior agrees well with the calculations of Ref.~\cite{Dondera_2012}, where similar deviations from the dipole prediction were reported for photon energies in 500 eV (cf. with Fig.~5 of Ref.~\cite{Dondera_2012}).

The corresponding photoelectron energy spectra are shown in Fig.~\ref{Fig 3}(b) for photon energies of 200~eV, 500~eV, and 1~keV. The open circles represent the calculations performed within the dipole approximation, while the solid curves denote the corresponding nondipole results obtained with the present Qprop-ND implementation. For all three photon energies, the first ATI peak constitutes the dominant contribution to the spectrum, whereas the higher-order ATI peaks decrease rapidly with increasing electron energy due to the progressively lower probability of absorbing multiple x-ray photons. As the photon energy increases from 200~eV to 1~keV, the overall ionization probability decreases by several orders of magnitude, resulting in a strong suppression of the higher-order ATI peaks. Remarkably, the dipole and nondipole spectra almost completely overlap for all photon energies considered, indicating that the leading-order nondipole interaction introduces only negligible corrections to the photoelectron energy spectra in this regime.
The present results are in excellent agreement with the benchmark calculations reported by Dondera and Bachau~\cite{Dondera_2012} (cf. with Fig.~4 of Ref.~\cite{Dondera_2012}),
providing a further validation of the accuracy and reliability of the present Qprop-ND implementation.

\subsection{Nondipole effect under linear and circular polarization}

\begin{figure}
    \centering
    \includegraphics[width=1\linewidth]{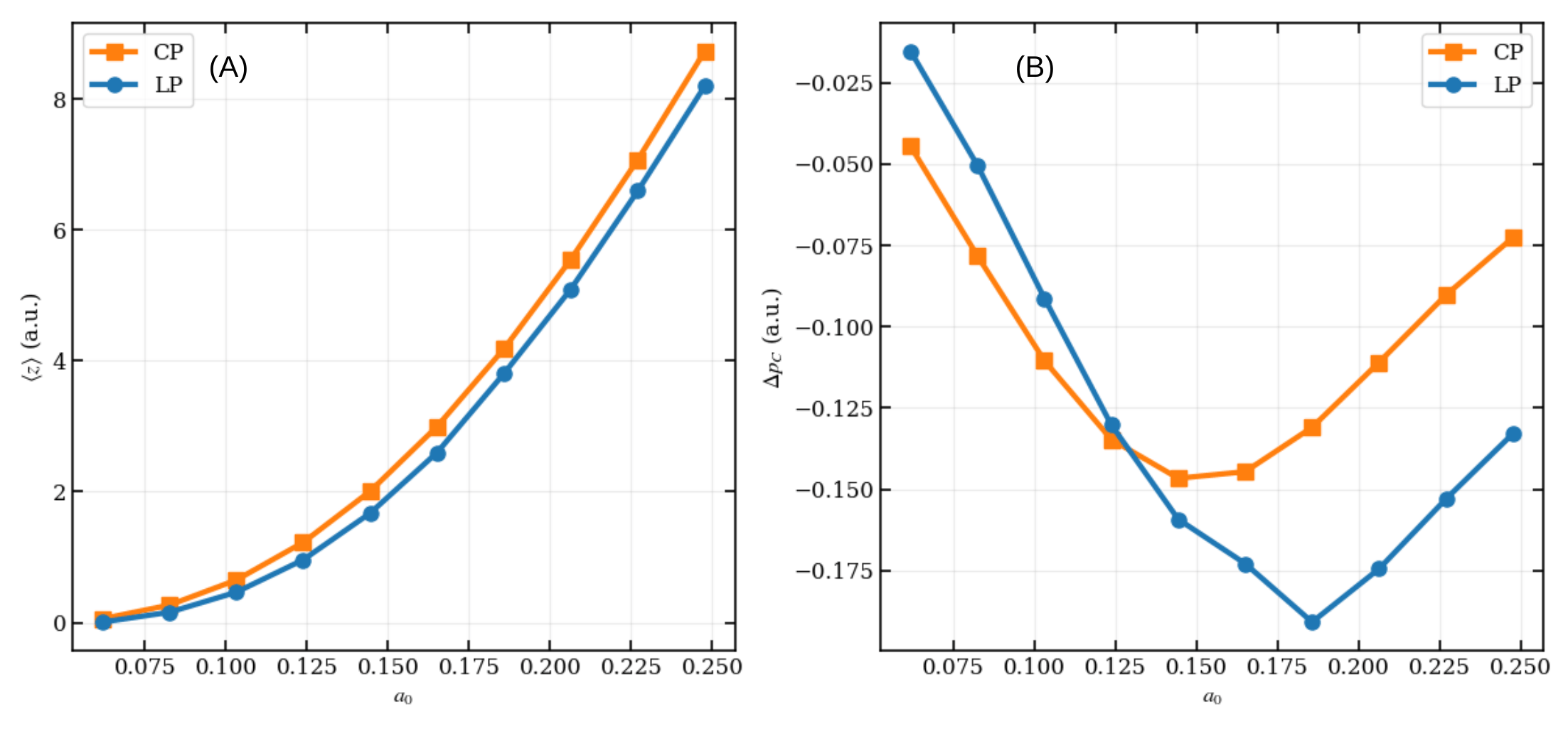}
    \caption{ZES for a hydrogen atom in a strong XUV field: (top row) the case of a linear polarization (LP) of the laser field, (bottom row) the case of a circular polarization (CP). The laser field for CP is \(E_{0}=40\), \(70\), and \(100\) a.u. in the left, middle, and right column, respectively. The corresponding LP fields satisfy the equal-intensity condition: $E_{0,\mathrm{lin}}=\sqrt{2}\,E_{0,\mathrm{circ}}$.
    The laser frequency is \(\omega = 5\) a.u., and laser pulses contain  \(n_{c}=10\) optical cycles.}
    \label{Fig 4}
\end{figure}

As an application of the developed Qprop-ND code, we next investigate the impact of the laser polarization on the zero-energy structure (ZES) in the photoelectron momentum distributions (PMDs).
The PMDs with a laser field of a linear and circular polarization  are shown in Fig.~\ref{Fig 4}. The employed vector potential for circular polarization is given by
\begin{align}
A_x(t) &= \frac{E_0}{\omega}
\sin^2\!\left(\frac{\omega t}{2n_c}\right)\sin(\omega t), \notag\\
A_y(t) &= \frac{E_0}{\omega}
\sin^2\!\left(\frac{\omega t}{2n_c}\right)\cos(\omega t). \notag
\end{align}
For the circularly polarized (CP) fields shown in the bottom row of Fig.~\ref{Fig 4}, the peak electric-field amplitudes are chosen as $E_{0}=40$, $70$, and $100$~a.u., respectively. The corresponding linearly polarized (LP) fields shown in the upper row satisfy the equal-intensity condition,
$E_{0,\mathrm{lin}}=\sqrt{2}\,E_{0,\mathrm{circ}}$, while the laser frequency and pulse duration are fixed at $\omega=5.0$~a.u. and $n_c=10$ optical cycles.

At relatively low laser intensities, the ZES exhibit qualitatively different pattern depending on the driving laser polarization.
For CP, the photoelectron emission is dominated by a single central lobe located along the laser counterpropagation direction. This lobe is a typical feature of the nondipole dynamics, originating from the asymmetry created by the  longitudinal nondipole drift accumulated during the laser pulse.
In contrast, for LP the two side lobes remain the dominant low-energy feature, while the central lobe is considerably weaker. This difference originates from the smaller peak electric field under the equal-intensity condition, which reduces the dipole-induced transverse momentum for CP (side lobes in ZES), whereas the leading-order nondipole momentum contribution remains of comparable magnitude for both polarization states. Consequently, the relative importance of nondipole effects is enhanced for CP.

\begin{figure}
    \centering
    \includegraphics[width=1\linewidth]{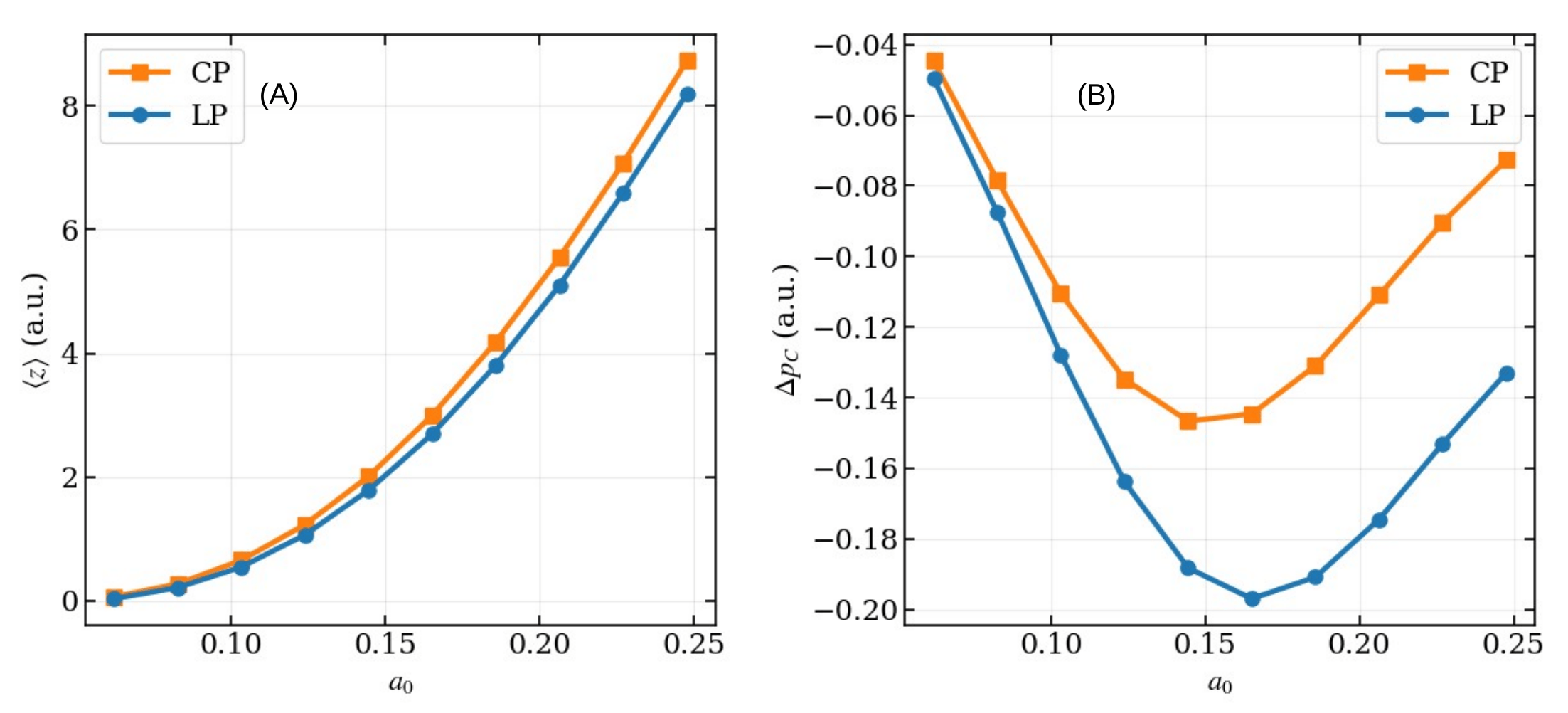}
    \caption{(A) Time evolution of the displacement expectation value $\langle z(t)\rangle$ for linear and circular polarization. (B) Corresponding accumulated Coulomb momentum transfer.}
    \label{Fig 5}
\end{figure}

As the laser intensity increases, additional interference lobes gradually emerge in both polarization configurations, giving rise to the characteristic three-lobe structure reported in previous nondipole studies. At still higher intensities, the PMDs evolve into petal-like patterns resembling Fresnel diffraction \cite{Geng_2021}. Although the overall topology becomes similar, clear polarization-dependent differences persist. LP produces broader and more elongated interference lobes in ZES with a pronounced asymmetry toward the laser counterpropagation direction, whereas CP yields more compact and nearly isotropic structures with significantly smoother angular modulation.

The distinct interference patterns originate from the different electron dynamics associated with the two polarization states. Under the equal-intensity condition, the peak electric field of a CP pulse is smaller, 
leading to a smaller classical excursion radius $\alpha_0=E_0/\omega^2$.
Since $\alpha_0$ determines the characteristic transverse extent of the emitted electron wavepacket, the smaller quiver radius in CP confines the spatial spreading of the wavepacket and consequently suppresses the formation of side lobes arising from nonadiabatic phase variations across the wavepacket.

An equally important contribution originates from the Coulomb interaction. Owing to the continuously rotating electric field, the ionized electron in a CP field is driven away from the ionic core and therefore experiences substantially weaker Coulomb attraction than in the LP case. This behavior is directly reflected in the expectation value of the longitudinal displacement $\langle z\rangle$ and of the accumulated Coulomb momentum transfer (CMT) $\Delta p_{zC}=-\langle\int_{t_i}^{t_f}\partial_z U(\mathbf{r})\rangle$, with the laser pulse switching-on and -off times $t_i$ and $t_f$, respectively, as shown in Fig.~\ref{Fig 5}. Throughout the investigated parameter range, the CP case consistently produces a larger forward displacement than the LP one because of the larger Coulomb attraction to the core in the case of LP hindering the nondipole drift. At large $a_0$ in Fig.~\ref{Fig 5}(b) the CMT is larger for LP as intuitively expected. The CMT in Fig.~\ref{Fig 5}(b) show a remarkable behaviour depending  on $a_0$: at low fields the CMT is increasing by module with larger $a_0$, while this behavior is reversed at large fields. Accordingly, we can call the first regime a Coulomb-influenced regime, while the second one -- the laser-dominated regime. 
In the CP case, the laser-driven longitudinal drift dominates the electron dynamics at considerably lower field strengths, causing the transition from the Coulomb-influenced regime to the laser-dominated regime to occur at a substantially smaller value of the normalized vector potential $a_0$ for CP than for LP.

\section{Conclusion}\label{concl}

We have upgraded the Qprop-3.0 code to a solution of the TDSE to include first-order ($\sim 1/c$) relativistic corrections. The developed Qprop-ND code is applicable for the investigation of nondipole regimes of the interaction of an electron with a laser field. The code is benchmarked against the available results of strong field ionization of a hydorgenlike atom in ultrastrong XUV and x-ray fields in the nondipole regime. The  results obtained with Qprop-ND are in good agreement with the existing ones, validating Qprop-ND for accuracy and reliability. 

As an application of the developed Qprop-ND code, we have investigated how the zero-energy structures in the photoelectron momentum distributions in ionization with an ultrastrong XUV field are modified when changing the laser polarization from LP to CP. We have identified two interaction regimes depending on the laser field strength: a Coulomb-dominated regime and the laser-dominated regime, showing that the transition to the laser-dominated regime occurs for CP at a  smaller value of the normalized vector potential $a_0$ than for LP.\\

\vskip 0.6cm

\appendix

{\Large \textbf{Appendix}}

\section{Clebsch--Gordan coefficients}\label{app_A}

We have used the following spherical harmonic relations:
\begin{equation}
\langle \ell m | Y_{1}^{0} Y_{1}^{\pm 1} | \ell' m' \rangle = \frac{3}{2\sqrt{15\pi}} \langle \ell m | Y_{2}^{\pm 1} | \ell' m' \rangle
\end{equation}
\begin{equation}
\langle \ell m | Y_{1}^{0} Y_{1}^{0} | \ell' m' \pm 1 \rangle = \frac{1}{\sqrt{5\pi}} \langle \ell m | Y_{2}^{0} | \ell' m' \pm 1 \rangle + \frac{1}{2\sqrt{\pi}} \langle \ell m | Y_{0}^{0} | \ell' m' \pm 1 \rangle
\end{equation}
Three spherical harmonics integrated over the solid angle $\Omega$ can be expressed in terms of Clebsch--Gordan coefficients 
$C^{c \, \gamma }_{a \alpha b \beta}$,
\begin{equation}
\langle \ell m | Y_{L}^{M} | \ell' m' \rangle
= \int d\Omega \, Y_{\ell}^{m^*} Y_{L}^{M} Y_{\ell'}^{m'}
= \sqrt{ \frac{ (2L + 1)(2\ell' + 1) }{ 4\pi (2\ell + 1) } } 
\, C_{\ell' 0 \, L 0}^{\ell 0} 
C_{\ell' m' \, L M}^{\ell m} 
\end{equation}
The Clebsch--Gordan coefficients arising from the nondipole terms in Eq.~(\ref{lm}) couple neighboring $\ell$s and $m$s are given by
\[
\begin{aligned}
\langle \ell m | Y_{2}^{1} | \ell' m' \rangle 
&= \frac{\sqrt{30\pi}}{{4\pi}} \, \delta_{m, m'+1} 
\Biggl(
  \delta_{\ell, \ell'+2} 
  \frac{1}{2\ell-1}\sqrt{ \frac{ (\ell + m - 2)(\ell + m-1)(\ell+m)(\ell-m) }{ (2\ell + 1)(2\ell - 3) } }
\\
&\quad
  - \delta_{\ell, \ell'-2} 
  \frac{1}{2\ell+3}\sqrt{ \frac{ (\ell + m + 1)(\ell - m + 1)(\ell -m + 2)(\ell -m + 3) }
  { (2\ell + 1)(2\ell + 5) } }
  \\
&\quad
  - \delta_{\ell,\ell'}
\frac{(1-2m)}{(2\ell-1)(2\ell+3)} 
\sqrt{ (\ell+m) (\ell-m+1)}
\Biggr) ,
\\[6pt]
\langle \ell m | Y_{2}^{-1} | \ell' m' \rangle 
&= \frac{\sqrt{30\pi}}{{4\pi}} \, \delta_{m, m'-1} 
\Biggl(
  \delta_{\ell, \ell'+2} 
  \frac{1}{2\ell-1}\sqrt{ \frac{ (\ell - m - 2)(\ell - m-1)(\ell+m)(\ell-m) }{ (2\ell + 1)(2\ell - 3) } }
\\
&\quad
  - \delta_{\ell, \ell'-2} 
  \frac{1}{2\ell+3}\sqrt{ \frac{ (\ell + m + 1)(\ell - m + 1)(\ell +m + 2)(\ell +m + 3) }
  { (2\ell + 1)(2\ell + 5) } }
  \\
&\quad
  - \delta_{\ell,\ell'}
\frac{(1+2m)}{(2\ell-1)(2\ell+3)} 
\sqrt{(\ell-m)(\ell+m+1)}
\Biggr) ,
\\[6pt]
\langle \ell m | Y_{2}^{0} | \ell' m' + 1 \rangle 
&= \frac{3\sqrt{5\pi}}{{4\pi}} \, \delta_{m, m'+1} 
\Biggl(
  \delta_{\ell, \ell'+2} 
  \frac{1}{2\ell-1}\sqrt{ \frac{ (\ell + m - 1)(\ell - m-1)(\ell+m)(\ell-m) }{ (2\ell + 1)(2\ell - 3) } }
\\
&\quad
  + \delta_{\ell, \ell'-2} 
  \frac{1}{2\ell+3}\sqrt{ \frac{ (\ell + m + 1)(\ell - m + 1)(\ell +m + 2)(\ell -m + 2) }
  { (2\ell + 1)(2\ell + 5) } }
\\
&\quad
 +\delta_{\ell, \ell'}
  \frac{2\bigl(\ell(\ell+1)-3m^2\bigr)}
             {3(2\ell-1)(2\ell+3)}
\Biggr) ,
\\[6pt]
\langle \ell m | Y_{2}^{0} | \ell' m' - 1 \rangle 
&= \frac{3\sqrt{5\pi}}{{4\pi}} \, \delta_{m, m'-1} 
\Biggl(
  \delta_{\ell, \ell'+2} 
  \frac{1}{2\ell-1}\sqrt{ \frac{ (\ell + m - 1)(\ell - m-1)(\ell+m)(\ell-m) }{ (2\ell + 1)(2\ell - 3) } }
\\
&\quad
  + \delta_{\ell, \ell'-2} 
  \frac{1}{2\ell+3}\sqrt{ \frac{ (\ell + m + 1)(\ell - m + 1)(\ell +m + 2)(\ell -m + 2) }
  { (2\ell + 1)(2\ell + 5) } }
  \\
&\quad
 +\delta_{\ell, \ell'}
  \frac{2\bigl(\ell(\ell+1)-3m^2\bigr)}
             {3(2\ell-1)(2\ell+3)}
\Biggr) .
\end{aligned}
\]

\section{Coefficients in the nondipole equation} \label{app_B}
The coefficients in Eqs.~(\ref{glm})-(\ref{elm}) are given by
 \[
g_{\ell m}
=
\frac{1+2m}{(2\ell-1)(2\ell+3)}
\sqrt{\frac{(\ell-m)(\ell+m+1)}{\ell(\ell+1)}}
\]

\[
h_{\ell m}
=
\frac{1}{2\ell+3}
\sqrt{
\frac{(\ell+m+1)(\ell-m+1)(\ell+m+2)(\ell+m+3)}
{(2\ell+1)(2\ell+5)}
}
\]

\[
f_{\ell m}
=
\frac{1}{2\ell+3}
\sqrt{
\frac{(\ell+m+1)(\ell-m+1)(\ell-m+2)(\ell-m+3)}
{(2\ell+1)(2\ell+5)}
}
\]

\[
p_{\ell m}
=
\sqrt{
\frac{(\ell+m+1)(\ell-m+1)}
{(2\ell+1)(2\ell+3)}
}
\]

\[
n_{\ell m}
=
\frac{m(1+2m)}{(2\ell-1)(2\ell+3)}
\sqrt{\frac{(\ell-m)(\ell+m+1)}{\ell(\ell+1)}}
+
\sqrt{\frac{\ell(\ell+1)-m(m+1)}{3}}
\]

\[
\tilde{e}_{\ell m}
=
\frac{1}{2\ell+3}
\sqrt{
\frac{(\ell+m+1)(\ell-m+1)(\ell+m+2)}
{(2\ell+1)(2\ell+5)}
}
\Big[
\left(m+\tfrac12\right)\sqrt{\ell+m+3}
+
\sqrt{(\ell-m+2)\big[(\ell+2)(\ell+3)-(m+1)m\big]}
\Big]
\]

\[
e_{\ell m}
=
\frac{1}{2\ell+3}
\sqrt{
\frac{(\ell+m+1)(\ell-m+1)(\ell-m+2)}
{(2\ell+1)(2\ell+5)}
}
\Big[
\left(m+\tfrac12\right)\sqrt{\ell-m+3}
-
\sqrt{(\ell+m+2)\big[(\ell+2)(\ell+3)-(m-1)m\big]}
\Big]
\]

\vskip 1cm
 
\bibliographystyle{elsarticle-num}

\bibliography{strong_fields_bibliography}

\end{document}